\RequirePackage{fix-cm}
\RequirePackage[colorlinks,citecolor=blue,urlcolor=blue,linkcolor=blue]{hyperref}
\documentclass[twocolumn]{svjour3}          

\smartqed  
\usepackage{graphicx}
\usepackage{booktabs}
\usepackage{subfigure}
\usepackage{placeins}

\usepackage{amsmath}
\usepackage{lineno}

\usepackage[british]{babel}
\usepackage[numbers,square,comma,sort&compress]{natbib}
\usepackage{xspace}
\providecommand{\HERAFitter}{{\texttt{HERA\-Fitter}}\xspace}

\journalname{DESY Report 15-035}

\begin{document}\sloppy

\title{QCD analysis of $W$- and $Z$-boson production at Tevatron}
\subtitle{ 
}
\author{\HERAFitter developers' team: 
S.~Camarda$^{1}$\and
P.~Belov$^{1,2}$\and
A.M.~Cooper-Sarkar$^{3}$\and 
C.~Diaconu$^{4}$\and
A.~Glazov$^{1}$\and
A.~Guffanti$^{5}$\and
A.~Jung$^{6}$\and
V.~Kolesnikov$^{7}$\and
K.~Lohwasser$^{8}$\and
V.~Myronenko$^{1}$\and
F.~Olness$^{9}$\and
H.~Pirumov$^{1}$\and
R.~Pla\v cakyt\. e$^{1}$\and
V.~Radescu$^{10}$\and
A.~Sapronov$^{7}$\and
W.~Slominski$^{11}$\and    
P.~Starovoitov$^{1}$\and    
M.~Sutton$^{12}$\and    
\vspace{0.5cm}
}


\institute{$^1$ Deutsches Elektronen-Synchrotron (DESY), Notkestrasse 85, D-22607 Hamburg, Germany\\
$^{2}$ Current address: Department of Physics, St. Petersburg State University, Ulyanovskaya 1, 198504 St. Petersburg, Russia\\
$^{3}$ Department of Physics, University of Oxford, Oxford, United Kingdom \\
$^{4}$ CPPM, IN2P3-CNRS, Univ. Mediterranee, Marseille, France \\
$^{5}$ Niels Bohr International Academy and Discovery Center, Niels Bohr Institute, University of Copenhagen, Blegdamsvej 17, DK-2100 Copenhagen, Denmark \\
$^{6}$ FERMILAB, Batavia, IL, 60510, USA \\
$^{7}$ Joint Institute for Nuclear Research (JINR), Joliot-Curie 6, 141980, Dubna, Moscow Region, Russia \\
$^{8}$ Deutsches Elektronen-Synchrotron (DESY), Platanenallee 6, D–15738 Zeuthen, Germany \\
$^{9}$ Southern Methodist University, Dallas, Texas \\
$^{10}$ Physikalisches Institut, Universit\"at Heidelberg, Heidelberg, Germany \\
$^{11}$ M. Smoluchowski Institute of Physics, Jagiellonian University, Cracow, Poland \\
$^{12}$ University of Sussex, Department of Physics and Astronomy, Sussex House, Brighton BN1 9RH, United Kingdom \\
%
}

\date{}

\maketitle

\begin{abstract}
Recent measurements of the $W$-boson charge asymmetry and of the
$Z$-boson production cross sections, performed at the Tevatron
collider in Run II by the D0 and CDF collaborations, are studied
using the HERAFitter framework to assess their impact on the 
proton parton distribution functions (PDFs).
The Tevatron measurements, together with deep-inelastic scattering
data from HERA, are included in a QCD analysis performed at
next-to-leading order, and compared to the predictions obtained using
other PDF sets from different groups. Good agreement between measurements
and theoretical predictions is observed. The Tevatron data
provide significant constraints on the $d$-valence quark distribution.

\end{abstract}

\setcounter{tocdepth}{2}

\section{Introduction}
\label{intro}
Accurate knowledge of the parton distribution functions (PDFs) is
essential for predictions at hadron colliders. The primary source
of information on the proton PDFs comes from deep-inelastic scattering
(DIS).
Measurements at fixed target experiments and at the HERA $e^{\pm} p$
collider provide constraints on the quark and gluon densities, and
discrimination of the quark flavours.
The DIS proton data mostly constrain the $u$-type quark density, due to the
greater couplings to the photon at low absolute four momentum
transfers, $Q^2$, whereas the $d$-type quark densities are only
constrained at high $Q^2$ with limited precision.
Even more challenging is the separation of the $d$-valence quark density
which relies on the HERA $e^+$ charge current data, which are
statistically limited in the published HERA I combined
data~\cite{Aaron:2009aa}. A better flavour separation is needed to
challenge the limits of precision physics at the LHC.

Drell-Yan production of $W$ and $Z$ bosons in proton-antiproton and
proton-proton collisions can provide additional information on the
$d$-quark PDFs. At leading order (LO) in QCD, the Drell-Yan processes
probe the PDFs at energy scales $Q$ corresponding to the boson masses, $m_V =
m_W$ and $m_V=m_Z$, and momentum fractions carried by the interacting
partons of $x_{1,2} = m_V/\sqrt{S} e^{\pm y}$, where $\sqrt{S}$ is
the centre-of-mass energy and $y$ is the boson rapidity. 

At the Tevatron proton-antiproton collider, the production of $W$ and $Z$
bosons is dominated by valence-quark interactions. The $Z$-boson
production has similar couplings for $u\bar{u}$ and $d\bar{d}$ fusion
processes, whereas $W$ bosons are produced predominantly by $u\bar{d}$
and $d\bar{u}$ fusions for $W^+$ and $W^-$ bosons, respectively.
Various measurements of $Z$-boson inclusive production and of
$W$-boson charge asymmetry have been reported by the D0 and CDF
collaborations~\cite{Abazov:2007jy,Aaltonen:2010zza,Abazov:2013rja,Aaltonen:2009ta,Abazov:2013dsa,D0:2014kma}. Some
of these data samples were included in previous PDF
studies~\cite{Gao:2013xoa,Harland-Lang:2014zoa,Ball:2014uwa,Owens:2012bv}. The
addition of the Tevatron data resulted in improved
PDFs, but some tensions were observed with global PDF
fits~\cite{Lai:2010vv,Catani:2010en,Thorne:2010kj,Ball:2010de}.

In this paper the data collected at the Tevatron collider in Run II
are analysed to assess their impact on the PDFs. The assumptions of
the correlation model of the experimental systematic uncertainties are
revised with respect to the recommendation of the experiments, leading
to improved agreement with the theoretical predictions. The analysis
is performed using the \HERAFitter
framework~\cite{Alekhin:2014irh,HERAFitter,Aaron:2009kv,Aaron:2009aa}
at next-to-leading order (NLO) QCD.
The Tevatron $W$- and $Z$-boson measurements are also compared to
predictions evaluated with the recent PDF sets CT10nlo~\cite{Gao:2013xoa},
MMHT2014~\cite{Harland-Lang:2014zoa} and
NNPDF3.0~\cite{Ball:2014uwa}. The impact of the Tevatron data on 
PDFs is studied using hessian
profiling~\cite{Paukkunen:2014zia} and Bayesian
reweighting~\cite{Giele:1998gw,Ball:2011gg,Ball:2010gb}
techniques. The profiling of PDF uncertainties is generalised to the
case of asymmetric PDF uncertainties.

This paper is organised as follows: the data samples are introduced in
Sec.~\ref{sec:data} and the theoretical predictions are discussed in
Sec.~\ref{sec:theorypred}. The QCD analysis settings and the methods
for comparing data with predictions based on existing PDFs are
discussed in Sec.~\ref{sec:theory}. Section~\ref{sec:results} reports
the results of the PDF analysis. The results obtained in the paper are
summarised in Sec.~\ref{sec:conclusions}.

\section{Experimental Measurements}
\label{sec:data}
\subsection{Data Sets}
The most recent measurements of $W$-boson charge asymmetry and
$Z$-boson inclusive production performed in Run II of the Tevatron
collider are considered in this study. They include the $Z$-boson
differential cross section as a function of rapidity, measured by the
D0 collaboration with $0.4$~fb$^{-1}$ of integrated luminosity in the
$Z \to ee$ channel~\cite{Abazov:2007jy}; the $Z$-boson differential
cross section as a function of rapidity, measured by the CDF
collaboration with $2.1$~fb$^{-1}$ of integrated luminosity in the $Z
\to ee$ channel~\cite{Aaltonen:2010zza}; the charge asymmetry of
muons as a function of rapidity in $W \to \mu \nu$ decays, measured by
the D0 collaboration with $7.3$~fb$^{-1}$ of integrated
luminosity~\cite{Abazov:2013rja}; the $W$-boson charge
asymmetry as a function of rapidity in the $W \to e \nu$ decay
channel, measured by the CDF collaboration with $1$~fb$^{-1}$ of
integrated luminosity~\cite{Aaltonen:2009ta}; the $W$-boson charge asymmetry as a
function of rapidity in the $W \to e \nu$ decay channel, measured by
the D0 collaboration with $9.7$~fb$^{-1}$ of integrated
luminosity~\cite{Abazov:2013dsa}. These measurements supersede 
the previous Run II Tevatron measurements of $W$-boson
charge asymmetry and $Z$-boson inclusive production.
Recently, the D0 collaboration has also released 
a measurement of the charge asymmetry of electrons as a function of
rapidity in $W \to e \nu$ decays~\cite{D0:2014kma}. However, this
measurement is performed with the same data set and event selection as
the measurement of Ref.~\cite{Abazov:2013dsa}, and cannot be included
simultaneously in a PDF fit without provision of the correlation
information. The Tevatron $W$- and $Z$-boson measurements considered
in this study are summarised in Tab.~\ref{tab:tevwzdata}.

Besides the Tevatron $W$- and $Z$-boson measurements,
the HERA I combined measurements of the inclusive 
DIS neutral- and charged-current 
cross sections measured by the H1 and ZEUS
experiments~\cite{Aaron:2009aa} are used in this study.
The neutral-current measurements cover a wide range in Bjorken-$x$ and $Q^2$, 
which is essential for the determination of PDFs,
whereas the charged-current measurements provide further information
to disentangle the contributions in PDFs from $u$-type and $d$-type
quarks and anti-quarks at $x>0.01$.
The DIS data are required to be in the kinematic region
$Q^2>Q^2_{\textrm{min}}=7.5$~GeV$^2$, where perturbative QCD
calculations are reliable.

\subsection{Experimental Uncertainties}
Statistical uncertainties are considered to be uncorrelated between bins,
with the exception of the D0 measurement of $W$-boson charge asymmetry
of Ref.~\cite{Abazov:2013dsa}, for which bin-to-bin statistical
correlations are provided.

In general, the correlation model of the experimental uncertainties
recommended by the Tevatron experiments is adapted and followed in the QCD
analysis, with the exception of the experimental systematic
uncertainties related to trigger and lepton identification efficiencies.
These uncertainties are provided by the D0 and CDF experiments in the
form of total uncertainties in each bin of the measurements. However,
the trigger and lepton identification corrections are
estimated from data, and they are influenced, among other effects, by
statistically uncorrelated bin-to-bin fluctuations. Since the exact
bin-to-bin correlation pattern of these uncertainties is not provided
by the experiments, a conservative approach is followed in this study,
and the uncertainties related to trigger and lepton identification
efficiencies are treated as uncorrelated bin-to-bin for the nominal
fit. According to this prescription, the following uncertainties 
are treated as uncorrelated bin-to-bin:
the central- and forward-electron identification efficiencies of
Ref.~\cite{Aaltonen:2010zza}, the trigger isolation efficiency of
Ref.~\cite{Abazov:2013rja}, the trigger and electron
identification efficiencies of Ref.~\cite{Aaltonen:2009ta}, and the
electron identification, charge misidentification and positron to
electron efficiency corrections of Ref.~\cite{D0:2014kma}.

All the other experimental systematic uncertainties are considered
fully correlated bin-to-bin, with the exception of the D0 measurement of
$W$-boson charge asymmetry of Ref.~\cite{Abazov:2013dsa}, where the
total experimental systematic uncertainty is treated as bin-to-bin
uncorrelated, as recommended by the D0 experiment, and for the
electron charge asymmetry in $W \to e \nu$ decays of
Ref.~\cite{D0:2014kma}, where the uncertainty of the unfolding
procedure due to the limited statistics of the Monte Carlo (MC) sample
is treated as uncorrelated bin-to-bin.
The dependence of the measured asymmetry on the PDF set used to
reconstruct the $W$-boson rapidity was studied in the D0
measurement of $W$-boson charge asymmetry of Ref.~\cite{Abazov:2013dsa}.
In this paper, the $W$-boson charge asymmetry extracted with
the CTEQ6.6 PDF set is used as the central value, and the $22$ CTEQ6.6
positive and negative PDF eigenvector variations are considered as
bin-to-bin correlated systematic uncertainties.

For the two measurements of $Z$-boson differential cross section as a
function of rapidity, statistical uncertainties are scaled to the
expected number of events assuming they are Poisson distributed.
The experimental systematic uncertainties are treated as
multiplicative, and linearly scaled to the expected cross sections,
except for the background uncertainties which are treated as additive,
and are not scaled.
For the measurements of $W$-boson and lepton charge asymmetry, all the
uncertainties are treated as additive, and are not scaled.

The statistical uncertainties of the HERA I data are treated as
uncorrelated and scaled to the expected number of events assuming
Poisson distribution, whereas the experimental systematic
uncertainties are fully correlated and are scaled linearly to the
expected cross sections.

\section{Theoretical Predictions}
\label{sec:theorypred}
The theoretical predictions corresponding to the Tevatron measurements
of $W$-boson charge asymmetry and $Z$-boson inclusive production are
included in the fits using APPLGRID~\cite{Carli:2005ji,Carli:2010rw} files.
These predictions have been evaluated with
MCFM~\cite{Campbell:1999ah,Campbell:2010ff} at NLO QCD according to
the phase-space definitions of each measurement, which are as
follows: the D0 and CDF measurements of the $Z$-boson differential
cross section as a function of rapidity are defined in the full
kinematic range of the decay leptons, without any requirements on the 
rapidity and $p_T$ of the leptons. In the D0 measurement, the invariant mass of the
dielectron system is defined in the range $71 < m_{ee} < 111$~GeV,
whereas in the CDF measurement, it is defined in the range $66 <
m_{ee} < 116$~GeV. The charge asymmetry of muons as
a function of rapidity in $W \to \mu \nu$ decays, and the charge
asymmetry of electrons in $W \to e \nu$ decays, measured by the D0
experiment, are defined with $p_T^{\ell} > 25$~GeV and $p_T^{\nu} > 25$~GeV.
The $W$-boson charge asymmetry as a function of
rapidity in the $W \to e \nu$ decay channel, measured by the CDF
collaboration, is defined in the full kinematic range, without any
requirements on the lepton rapidity and $p_T$. The corresponding D0
measurement of the $W$-boson charge asymmetry in the $W \to e \nu$ decay
channel is defined in a kinematic region where the charged lepton and
the neutrino are required to have $p_T> 25$~GeV without further
requirements on the lepton rapidity.
The kinematic requirements of the Tevatron $W$- and $Z$-boson
measurements are summarised in Tab.~\ref{tab:tevwzdata}.
Notice that the CDF and D0 measurements of $W$-boson charge asymmetry
in the $W \to e \nu$ decay channel of Refs.~\cite{Aaltonen:2009ta}
and~\cite{Abazov:2013dsa} are defined in different kinematic regions
and they should not be compared without extrapolating them to a common
phase space.
Tables of the Tevatron measurements, with updated correlation model,
and corresponding APPLGRID theoretical predictions are publicly
available at \emph{herafitter.org}.

The QCD predictions for the DIS cross sections are evaluated by solving 
the DGLAP evolution equations~\cite{Gribov:1972ri,Altarelli:1977zs,Curci:1980uw,Furmanski:1980cm,Moch:2004pa,Vogt:2004mw} 
at NLO in the $\overline{MS}$ scheme~\cite{PhysRevD.8.3497} using the QCDNUM 
program~\cite{Botje:2010ay} with the renormalisation and
factorisation scales set to $Q^2$.
The light quark coefficient functions are calculated in QCDNUM.
The heavy $c$- and $b$-quark distributions are dynamically generated, and the
corresponding coefficient functions for the neutral-current processes
with $\gamma^*$ exchange are calculated in the ge\-ne\-ral-mass
variable-flavour-number
scheme~\cite{Thorne:1997ga,Thorne:2006qt,Thorne:2012az}, with up to
five active quark flavours.
For the charged-current processes and the neutral-current processes
with a $Z$ contribution, the heavy quarks are treated as massless. 

\section{QCD Analysis}
\label{sec:theory}

\subsection{PDF Fit Settings}

\label{sec:pdffit}
The QCD analysis and PDF extraction is performed with the open-source
framework \HERAFitter.
The charm mass is set to $m_c = 1.38$~GeV, 
as estimated from HERA charm production cross section~\cite{Abramowicz:1900rp}, 
and the bottom mass to $m_b = 4.75$~GeV~\cite{Martin:2009iq}. 
The strong-interaction coupling constant at the $Z$ boson mass,
$\alpha_s(M_Z)$, is set to 0.118, and two-loop order is used for the running of $\alpha_s$.
%
%

The PDFs for the gluon, $u$-valence, $d$-valence, $\bar{u}$, $\bar{d}$
quark densities are parametrised at the input scale
$Q^2_0=1.7$~GeV$^2$ as follows:

\begin{eqnarray}
\label{eq:pdf} xf(x) = A_f x^{B_f} (1-x)^{C_f} (1 + D_fx + E_fx^2) e^{F_fx} \,;\\
\nonumber f = u_v, d_v, g, \bar{u}, \bar{d} \,.
\end{eqnarray}
The contribution of the $s$-quark density is taken to be proportional to the
$\bar{d}$-quark density by setting $x\bar{s}(x) = r_s x\bar{d}(x)$,
with $r_s=1.0$, as suggested in Ref.~\cite{Aad:2012sb}. The strange and
anti-strange quark densities are taken to be equal: $x\bar{s}(x) = x
s(x)$. The normalisation of the $x u_v(x)$ ($x d_v(x)$)
va\-lence-quark density, $A_{u_v}$ ($A_{d_v}$), is determined by the
quark-counting sum rule, whereas the normalisation of the gluon
density, $A_g$, is determined by the momentum sum rule. The $x\to0$
limit of the $u$- to $d$- sea quark densities is fixed to unity by
setting $B_{\bar{u}} = B_{\bar{d}}$ and $A_{\bar{u}} = A_{\bar{d}}$.

A $\chi^2$ function used for the data to theory comparison is defined
as in Ref.~\cite{Aaron:2009aa}, with an additional penalty term as
described in Ref.~\cite{Aaron:2012qi}, and minimised with
MINUIT~\cite{minuit} to extract the PDFs from the data.

\subsection{PDF Profiling}
The impact of a new data set on a given PDF set can be quantitatively
estimated with a profiling procedure~\cite{Paukkunen:2014zia}. The
profiling is performed using a $\chi^2$ function which includes both
the experimental uncertainties and the theoretical uncertainties
arising from PDF variations:

\begin{eqnarray}
\nonumber \lefteqn{\chi^2(\boldsymbol{\beta_{\rm exp}},\boldsymbol{\beta_{\rm th}}) = }  \\ 
&& \nonumber \sum_{i=1}^{N_{\rm data}} \frac{\textstyle \left( \sigma^{\rm exp}_i + \sum_j \Gamma^{\rm exp}_{ij} \beta_{j,\rm exp} - \sigma^{\rm th}_i - \sum_k \Gamma^{\rm th}_{ik}\beta_{k,\rm th} \right)^2}{\Delta_i^2} \\
&&   + \sum_j \beta_{j,\rm exp}^2 + \sum_k \beta_{k,\rm th}^2\,.   \label{eq:chi2prof}
\end{eqnarray}
The correlated experimental and theoretical uncertainties are included
using the nuisance parameter vectors $\boldsymbol{\beta_{\rm exp}}$
and $\boldsymbol{\beta_{\rm th}}$, respectively. Their influence on
the data and theory predictions is described by the $\Gamma^{\rm
exp}_{ij}$ and $\Gamma^{\rm th}_{ik}$ matrices. The index $i$ runs
over all $N_{\rm data}$ data points, whereas the index $j$ ($k$)
corresponds to the experimental (theoretical) uncertainty nuisance
parameters. The measurements and the uncorrelated experimental
uncertainties are given by $\sigma^{\rm exp}_i$ and $\Delta_i$\,, respectively, and
the theory predictions are $\sigma_i^{\rm th}$. The $\chi^2$ function of
Eq.~\ref{eq:chi2prof} can be generalised to account for asymmetric PDF
uncertainties:

\begin{eqnarray}
   \Gamma^{\rm th}_{ik} \to \Gamma^{\rm th}_{ik} +  \Omega^{\rm th}_{ik}\beta_{k, \rm th}\,, \label{eq:iter}
\end{eqnarray}
where $\Gamma^{\rm th}_{ik} = 0.5(\Gamma^{\rm th+}_{ik} - \Gamma^{\rm
th-}_{ik})$ and $\Omega^{\rm th}_{ik} = 0.5(\Gamma^{\rm th+}_{ik}
+ \Gamma^{\rm th-}_{ik})$ are determined from the shifts of
predictions corresponding to up ($\Gamma^{\rm th+}_{ik}$) and down
($ \Gamma^{\rm th-}_{ik}$) PDF uncertainty eigenvectors.

The minimisation of Eq.~\ref{eq:chi2prof} in its original form leads
to a system of linear equations. The generalised function, with
asymmetric PDF uncertainties, is minimised iteratively: the values of $\Gamma^{\rm
th+}_{ik}$ are updated using $\beta_{k, \rm th}$ from the
previous iteration and following the substitution of Eq.~\ref{eq:iter}.
Several iterations are required to converge, and the procedure is verified
using the MINUIT program which yields identical results.

The value at the minimum of the $\chi^2$ function provides a
compatibility test of the data and theory.
In addition, the values at the minimum of the nuisance parameters
$\beta^{\rm min}_{k,\rm th}$ can be interpreted as optimisation
(``profiling'') of PDFs to describe the data~\cite{Paukkunen:2014zia}. Explicitly, the profiled
central PDF set $f'_0$ is given by

\begin{equation}
   f'_0 = f_0 + \sum_k  \beta^{\rm min}_{k, {\rm
   th}} \left( \frac{f^{+}_k - f^{-}_k}{2}  - \beta^{\rm min}_{k, {\rm th}}  \frac{f^{+}_k + f^{-}_k - 2f_0}{2} \right)\,, \label{eq:prof}
\end{equation}
where $f_0$ is the original central PDF set and $f^{\pm}_k$ represents the
eigenvector sets corresponding to up and down variations.

The shifted PDFs have reduced uncertainties. In general, the
shifted eigenvectors are no longer orthogonal, but can be transformed
to an orthogonal representation using a standard diagonalisation
procedure, as in Ref.~\cite{Aaron:2009bp}.
In this method the covariance matrix $C$ of the PDF nuisance
parameters is diagonalised as

\begin{eqnarray}
  \lefteqn{\boldsymbol{\beta_{\rm th}^T} C \boldsymbol{\beta_{\rm th}} = \boldsymbol{\beta_{\rm th}^T} G^T D G \boldsymbol{\beta_{\rm th}} = 
  \boldsymbol{\beta_{\rm th}^T} (\sqrt{D} G)^T \sqrt{D}G \boldsymbol{\beta_{\rm th}}}  \nonumber \\
&&  =  (G' \boldsymbol{\beta_{\rm th}})^T  G'\boldsymbol{\beta_{\rm th}} = \boldsymbol{(\beta'_{\rm th})^T}  \boldsymbol{\beta'_{\rm th}}\,,
\label{eq:diag}
\end{eqnarray} 
where $G$ is an orthogonal matrix, $D$ is a positive definite diagonal
matrix, and $\sqrt{D}$ is a diagonal matrix built of
$\sqrt{D_{ii}}$. The matrices $G$ and $D$ can be constructed using
the eigenvectors and eigenvalues of the matrix $C$. The transformation
$G'$ can be adjusted, using orthogonal transformations, to keep the
new eigenvector basis aligned along the original as much as
possible. As a result of this adjustment, the transformation matrix
can take a triangular form with all diagonal elements greater than
zero.

The method can be extended to PDF sets with asymmetric
uncertainties: the transformation matrix is
determined using symmetrised uncertainties as in
Eq.~\ref{eq:diag}, and the orthogonal up and down PDF eigenvectors
$f^{+'}_i$ and $f^{-'}_i$ are calculated as

\begin{small}
\begin{eqnarray}
\nonumber   f^{+'}_i =& f'_0 + \sum_j G'_{ji} \left( \frac{f^{+}_j - f^{-}_j}{2} + G'_{ji} \frac{f^{+}_j + f^{-}_j - 2 f_0}{2}\right),  \\
\nonumber   f^{-'}_i =& f'_0 - \sum_j G'_{ji} \left( \frac{f^{+}_j - f^{-}_j}{2} - G'_{ji} \frac{f^{+}_j + f^{-}_j - 2 f_0}{2}\right).  
\end{eqnarray} 
\end{small}

\subsection{Bayesian Reweighting}
An alternative approach to assess the impact of new data on PDFs
is the Bayesian reweighting technique, first proposed
in Ref.~\cite{Giele:1998gw} and further developed by the NNPDF
collaboration~\cite{Ball:2011gg,Ball:2010gb}. The Bayesian reweighting
can be applied to PDF sets provided in the form of MC replicas, such
as the NNPDF3.0 set~\cite{Ball:2014uwa}.
Recently, a variant of the method which can be used with PDFs provided
in the eigenvector representation has been
developed~\cite{Watt:2012tq} and is also available in \HERAFitter.

The Bayesian reweighting is based on the assumption that an ensemble
of MC replicas provides a representation of the probability
distribution in the space of PDFs. For a given PDF set with $N_{\rm
rep}$ replicas $\{f_k\}$, with $k=1,2,...,N_{\mathrm{rep}}$, the
central value for a general observable,
$\mathcal{O}(\{f_k\})$, is estimated as the average of the
predictions obtained from the ensemble:

\begin{eqnarray}
\langle\mathcal{O}\rangle =  \frac{1}{N_{\mathrm{rep}}} \sum_{k=1}^{N_{\mathrm{rep}}} \mathcal{O}(f_{k})\,.
\end{eqnarray}

With the inclusion of new data, the probability distribution
associated with the original PDF set is modified according to Bayes
Theorem. For each replica $k$, a weight $w_k$ is obtained from the
$\chi^2$ function according to:

\begin{eqnarray}
 w_k = \frac{(\chi^2_k)^{\frac{1}{2} (N_{\mathrm{data}}-1) } e^{-\frac{1}{2}\chi^2_k}}
          { \frac{1}{N_{\mathrm{rep}}} \sum^{N_{\mathrm{rep}}}_{k=1}(\chi^2_k)^{\frac{1}{2}(N_{\mathrm{data}}-1)} e^{-\frac{1}{2}\chi^2_k}  }\,,
\end{eqnarray}
where $N_{\mathrm{data}}$ is the number of new data points and
$\chi^2_k$ is the $\chi^2$ value between data and predictions
corresponding to the $k$-th PDF replica.

The prediction for a given observable, after the inclusion of the new
data, is evaluated as the weighted average of predictions obtained
from the ensemble:

\begin{eqnarray} \label{eq:weighted}
\langle\mathcal{O}\rangle =  \frac{1}{N_{\mathrm{rep}}} \sum_{k=1}^{N_{\mathrm{rep}}} w_k \mathcal{O}(f_{k})\,.
\end{eqnarray}

The reweighting procedure is very fast and results in a new,
updated, MC PDF set. Some of the replicas of the PDF set may have very
small weights (typically those which do not describe the new data),
and they do not contribute to the ensemble any longer. The number of
{\em effective replicas}, $N_\mathrm{eff}$, of a reweighted set is
quantified by the Shannon entropy

\begin{eqnarray}
\label{eq:shannon}
N_\mathrm{eff}\equiv 
\exp\left\{\frac{1}{N_\mathrm{rep}}\sum_{k=1}^{N_\mathrm{rep}}w_k\ln(N_\mathrm{rep}/w_k)\right\}\,.
\end{eqnarray}
An un-weighting procedure can be performed on the MC set such that
PDFs with small weights are suppressed and a new set is produced,
which has unit weight for all PDF replicas in addition to
statistically reproducing the averages from Eq.~\ref{eq:weighted}.

\section{Results}
\label{sec:results}
The QCD fit analysis described in Sec.~\ref{sec:theory} is performed
on the Tevatron $W$- and $Z$-boson data, together with the HERA I
data. The fit is used to study the compatibility of the data with NLO
QCD predictions, and to assess the impact of the Tevatron data on
PDFs. The profiling and the reweighting techniques are used to asses
the impact of the Tevatron data on various PDF sets.

The optimal parametrisation for the PDF fit is found through
a \emph{parametrisation scan}, a procedure first introduced in
Ref.~\cite{Aaron:2009aa}. The scan is performed by starting from
a parametrisation with a basic polynomial form, where $D_f$, $E_f$,
and the exponential parameters $F_f$ of Eq.~\ref{eq:pdf} are set to zero.
After application of the quark-counting and momentum sum rules, and of
the $x \to 0$ constraints on $\bar{u}$ and $\bar{d}$, the initial PDFs
parametrisation has 10 free parameters.
The 15 $D_f$, $E_f$ and $F_f$ additional parameters are allowed to
vary, one parameter at a time, and the parameter which induces the
largest reduction of $\chi^2_{\textrm{min}}$ is added as a free
parameter for the next iteration of the scan.
The PDF fits which lead to solutions with negative high-$x$ PDFs
are discarded. For each PDF, the exponential term $e^{F_fx}$ and the polynomial term $(1 + D_fx + E_fx^2)$ 
are considered as mutually exclusive, that is, when the exponential term is preferred, the polynomial
term is no longer considered in the scan, and vice versa.
The procedure is stopped when
the reduction in the $\chi^2_{\textrm{min}}$ value, $\Delta \chi^2_{\textrm{min}}$, is less than unity.

Table~\ref{tab:parscan} shows the results of the parametrisation
study: the parameters which induce the largest $\Delta
\chi^2_{\textrm{min}}$ are, in order, $F_{d_v}$, $F_{u_v}$, $D_g$,
$D_{\bar{d}}$, and $D_{\bar{u}}$.
The optimal parametrisation found with this procedure has 15 free
parameters, and the PDFs are expressed as:
\begin{eqnarray}
  xg(x) &=& A_g x^{B_g} (1-x)^{C_g} (1 + D_gx)\,;  \label{eq:2} \\
  x u_v(x) &=& A_{u_v} x^{B_{u_v}}(1-x)^{C_{u_v}} e^{F_{u_v}x}\,;  \label{eq:3} \\
  x d_v(x) &=& A_{d_v} x^{B_{d_v}}(1-x)^{C_{d_v}} e^{F_{d_v}x}\,;  \label{eq:4} \\
  x\bar{u}(x) &=& A_{\bar{u}} x^{B_{\bar{u}}}(1-x)^{C_{\bar{u}}}(1 + D_{\bar{u}}x)\,; \label{eq:5} \\
  x\bar{d}(x) &=& A_{\bar{d}} x^{B_{\bar{d}}}(1-x)^{C_{\bar{d}}}(1 + D_{\bar{d}}x)\,.  \label{eq:6} 
\end{eqnarray}

The parametrisation of Eqs.~(\ref{eq:2}-\ref{eq:6}) is used for a fit to
the HERA I data, and for a combined fit to the HERA I and Tevatron $W$- and
$Z$-boson data. Table~\ref{tab:chi2fit} shows the
$\chi^2_{\textrm{min}}$ per degrees of freedom (dof) of the two fits. The contribution to the
total $\chi^2_{\textrm{min}}$ of each data set, henceforth referred
to as \emph{partial} $\chi^2$, is also shown.
The inclusion of the Tevatron $W$- and $Z$-boson data in the fit,
which corresponds to 93 additional points, results in an increase of about 110
in the overall $\chi^2_{\textrm{min}}$ of the fit, and the
partial $\chi^2$ per number of points of each of the Tevatron and HERA I
data set is close to unity.

\begin{figure*}
  \begin{center}
    \subfigure[]{\includegraphics[width=0.49\textwidth]{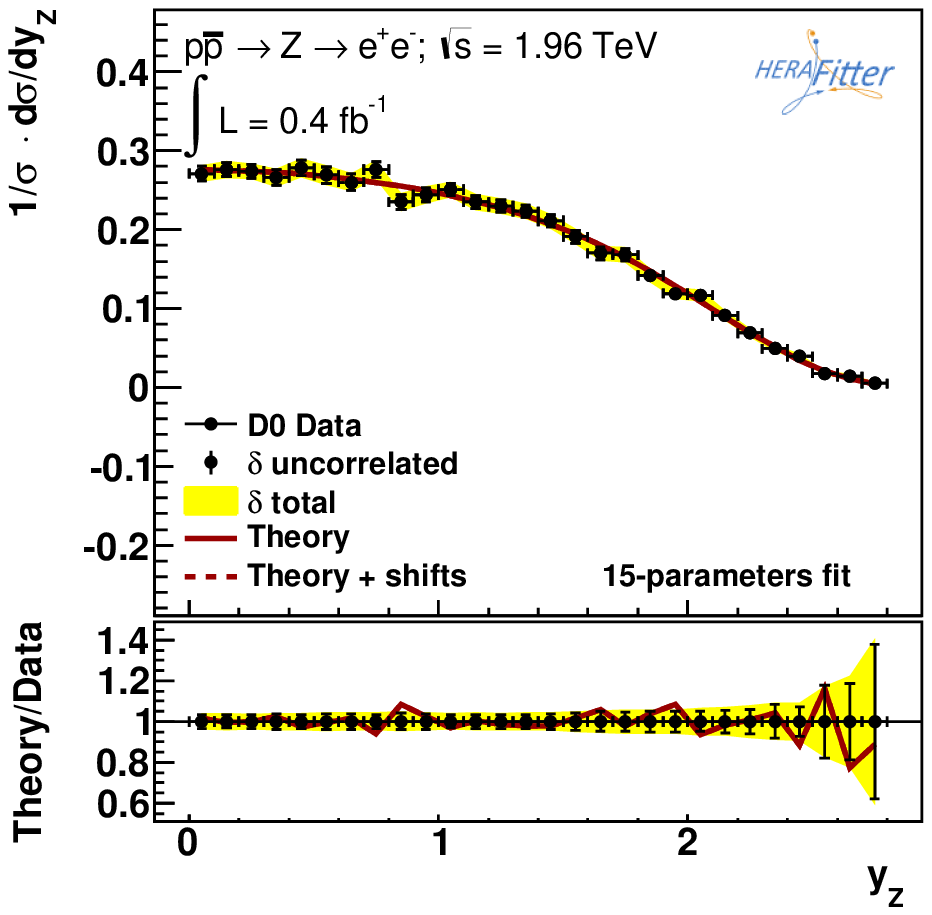}}
    \subfigure[]{\includegraphics[width=0.49\textwidth]{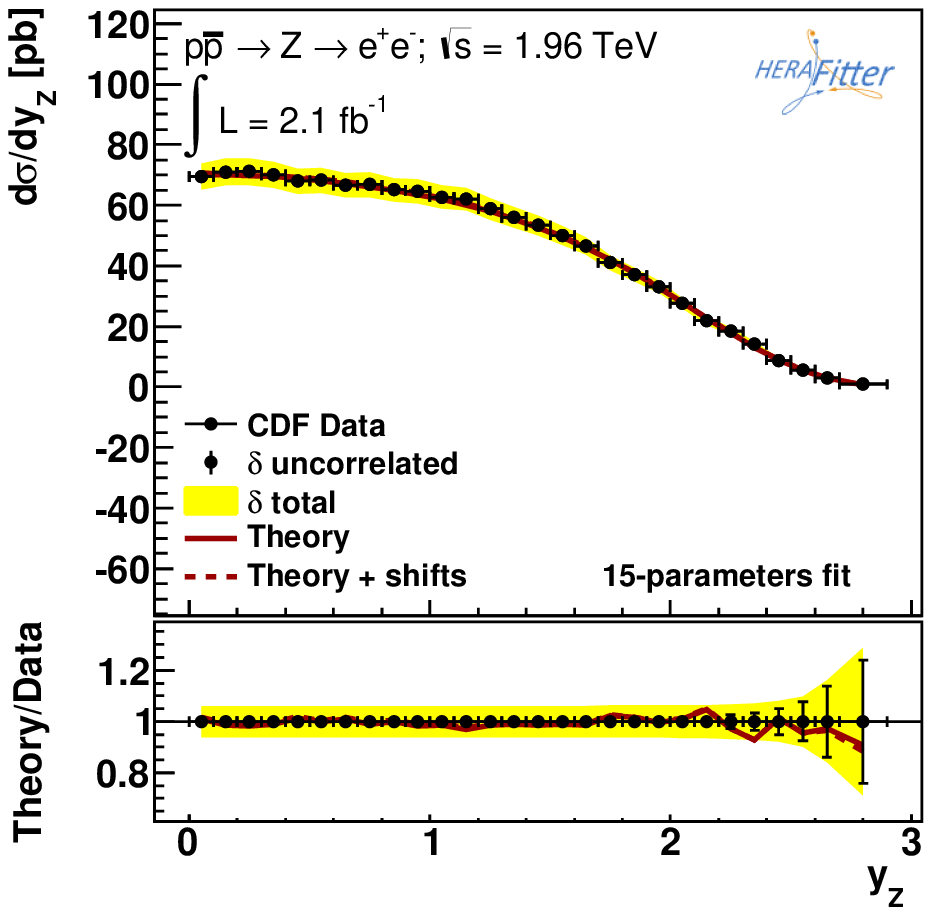}}\\
  \end{center}
  \caption{\label{fig:Zdatafit} Theoretical predictions evaluated
    with the PDFs extracted from a fit to the HERA I and
    Tevatron data are compared to (a) $Z$-boson differential cross section as a
    function of rapidity, measured by the D0 collaboration and (b)
    $Z$-boson differential cross section as a function of rapidity,
    measured by the CDF collaboration.
    The red continuous lines correspond to the theoretical
    predictions, the red dashed lines are the theoretical predictions
    shifted by the experimental shift terms $\sum_j \Gamma^{\rm
    exp}_{ij} \beta_{j,\rm exp}$ of Eq.~(\ref{eq:chi2prof}). The
    yellow bands show the total experimental uncertainty, the black
    vertical bars show the quadratic sum of statistical and
    systematic uncorrelated uncertainties. Note, the theoretical
    predictions and the theoretical predictions
    shifted by the experimental shift terms are nearly identical, and
    virtually indistinguishable in these plots.
    }
\end{figure*}

\begin{figure*}
    \begin{center}
    \subfigure[]{\includegraphics[width=0.49\textwidth]{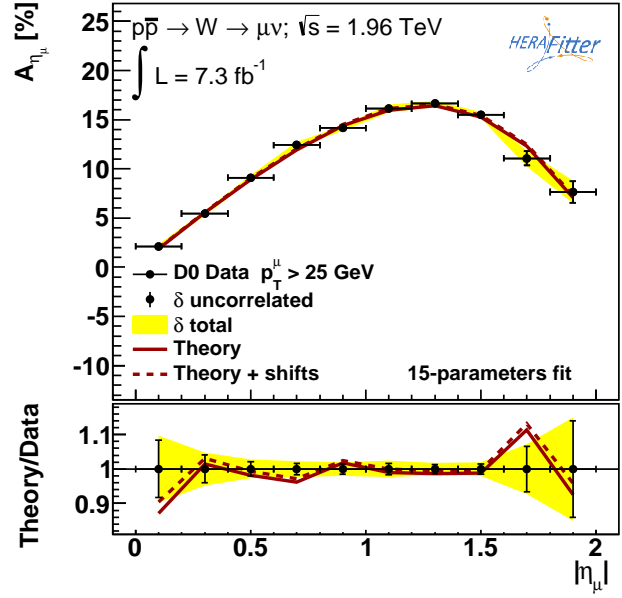}}\\
    \subfigure[]{\includegraphics[width=0.49\textwidth]{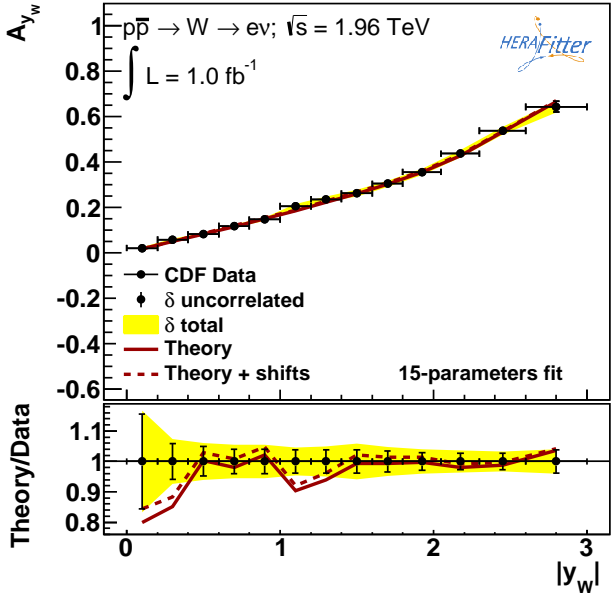}}
    \subfigure[]{\includegraphics[width=0.49\textwidth]{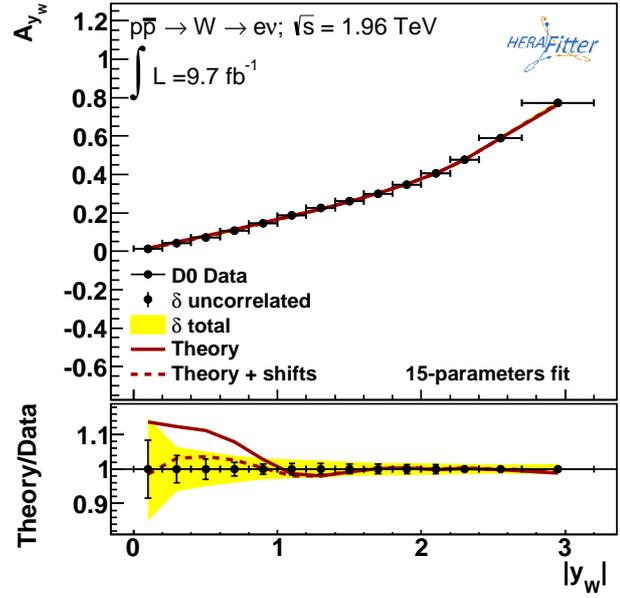}}
  \end{center}
  \caption{\label{fig:Wdatafit} Theoretical predictions evaluated
    with the PDFs extracted from a fit to the HERA I and
    Tevatron data are compared to (a) the charge asymmetry of muons as a
    function of rapidity in $W \to \mu \nu$ decays, measured by the D0
    collaboration, (b) the $W$-boson charge asymmetry as a
    function of rapidity in the $W \to e \nu$ decay channel, measured by
    the CDF collaboration, and (c) the $W$-boson charge asymmetry as a
    function of rapidity in the $W \to e \nu$ decay channel, measured by
    the D0 collaboration.
    The red continuous lines correspond to the theoretical
    predictions, the red dashed lines are the theoretical predictions
    shifted by the experimental shift terms $\sum_j \Gamma^{\rm
    exp}_{ij} \beta_{j,\rm exp}$ of Eq.~(\ref{eq:chi2prof}). The
    yellow bands show the total experimental uncertainty, the black
    vertical bars show the quadratic sum of statistical and
    systematic uncorrelated uncertainties.
}
\end{figure*}

Figures~\ref{fig:Zdatafit} and~\ref{fig:Wdatafit} show the Tevatron
$Z$- and $W$-boson measurements, respectively, compared to the
theoretical predictions evaluated with the PDFs extracted from the
combined fit to the HERA I and Tevatron data.

\begin{figure*}
    \begin{center}
    \subfigure[]{\includegraphics[width=0.49\textwidth]{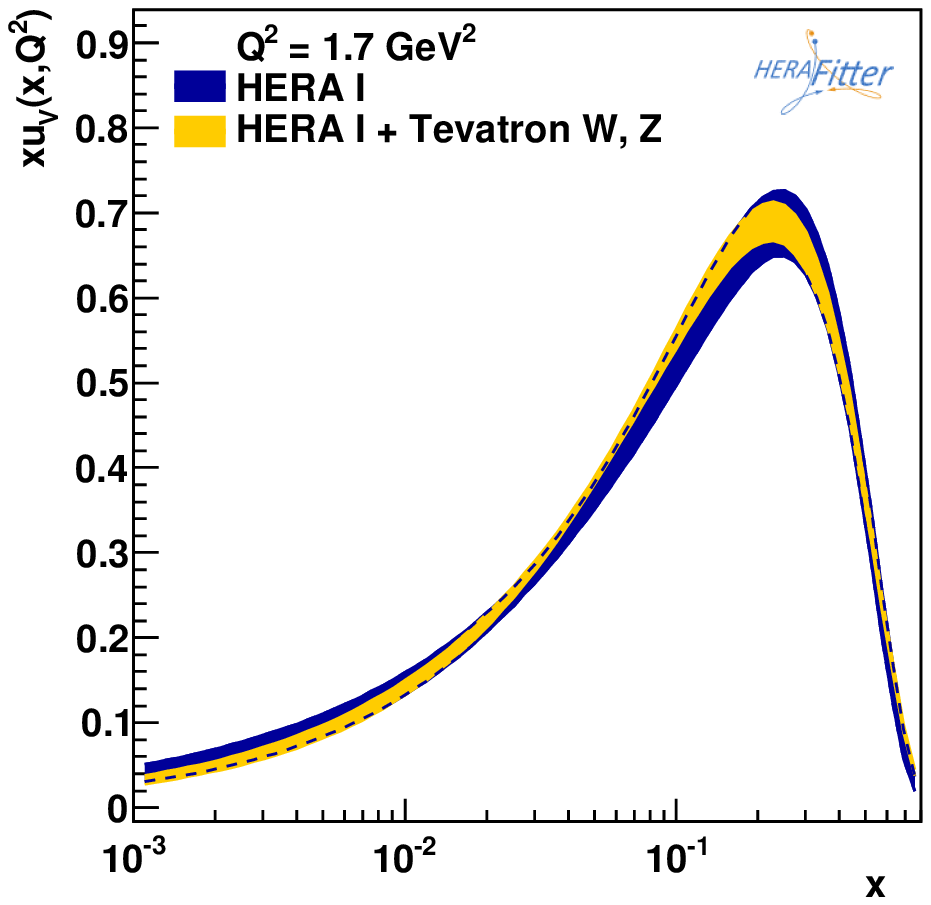}}
    \subfigure[]{\includegraphics[width=0.49\textwidth]{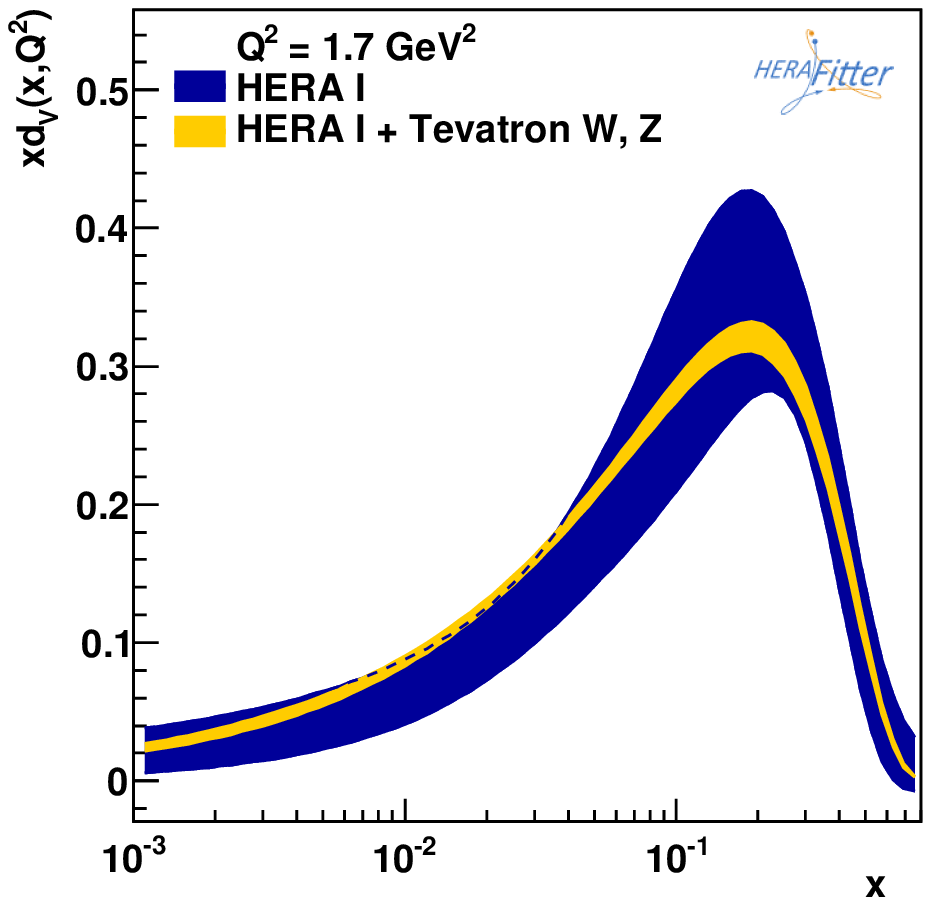}}
    \subfigure[]{\includegraphics[width=0.49\textwidth]{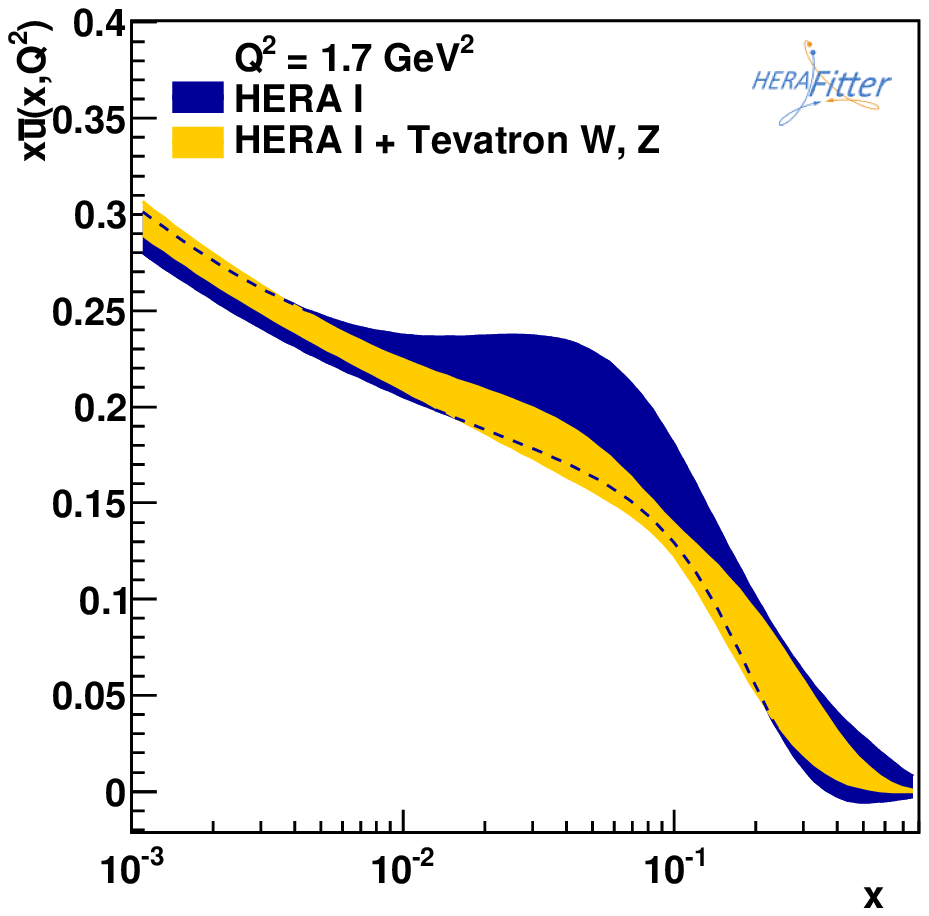}}
    \subfigure[]{\includegraphics[width=0.49\textwidth]{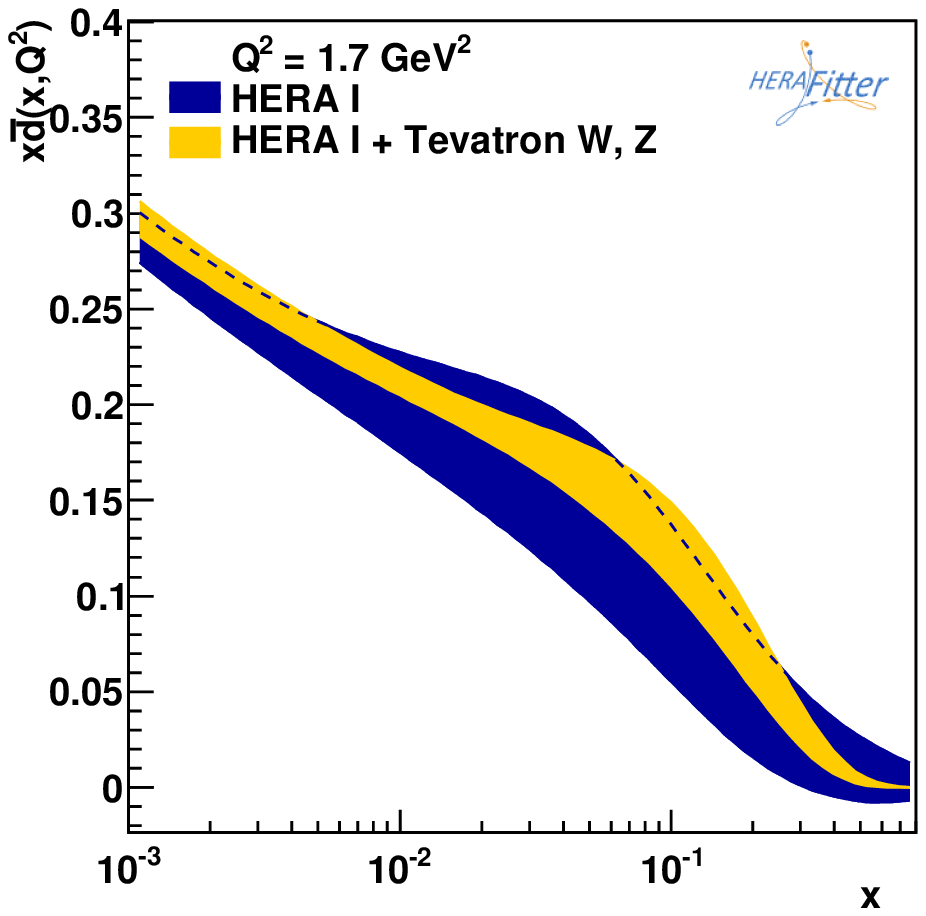}}
  \end{center}
  \caption{\label{fig:pdffit} PDFs at the starting scale $Q^2
    =1.7$~GeV$^2$ as a function of Bjorken-$x$ for (a) $u_v$, (b)
    $d_v$, (c) $\bar{u}$, and (d) $\bar{d}$, determined with a fit
    to the HERA I data (blue), and with a fit to the HERA I and
    Tevatron $W$- and $Z$-boson data (yellow).}
\end{figure*}

\begin{figure*}
  \begin{center}
    \subfigure[]{\includegraphics[width=0.49\textwidth]{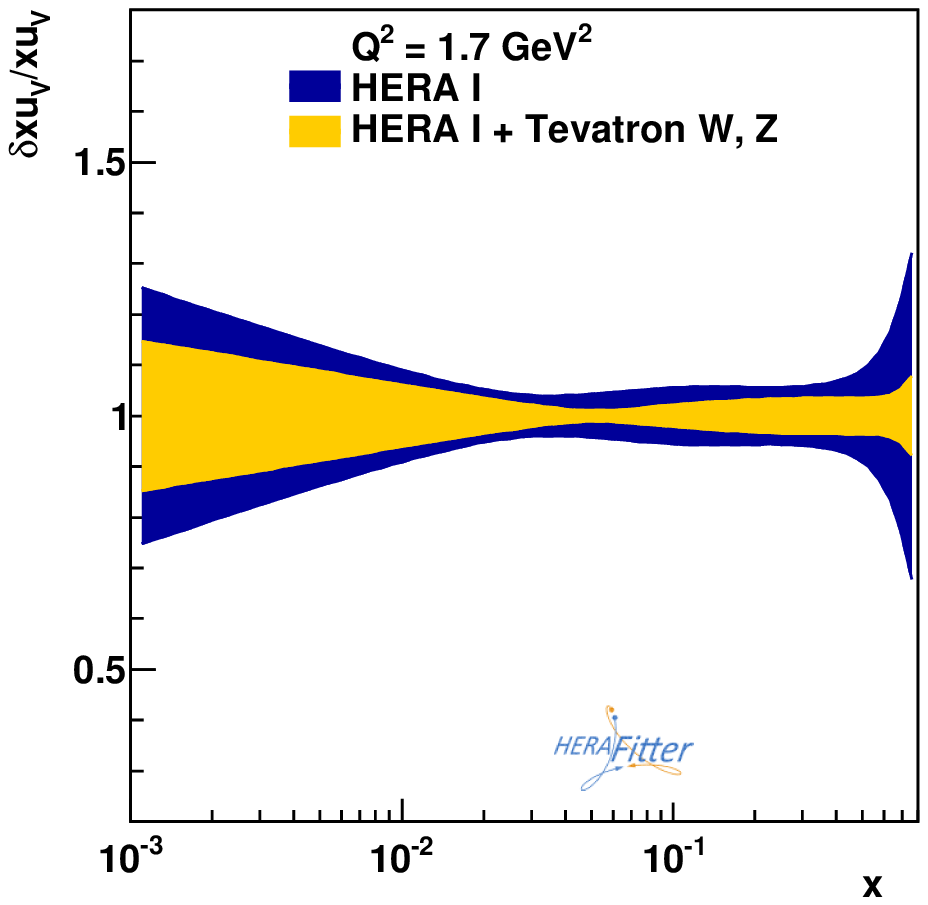}}
    \subfigure[]{\includegraphics[width=0.49\textwidth]{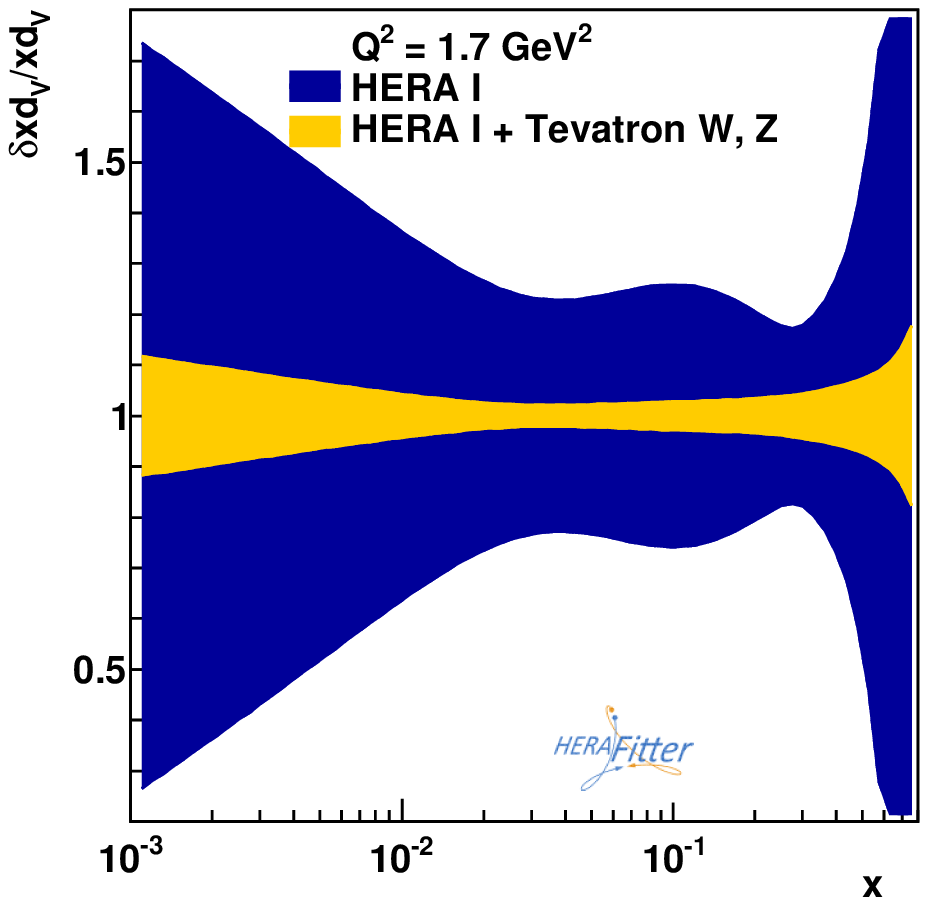}}
    \subfigure[]{\includegraphics[width=0.49\textwidth]{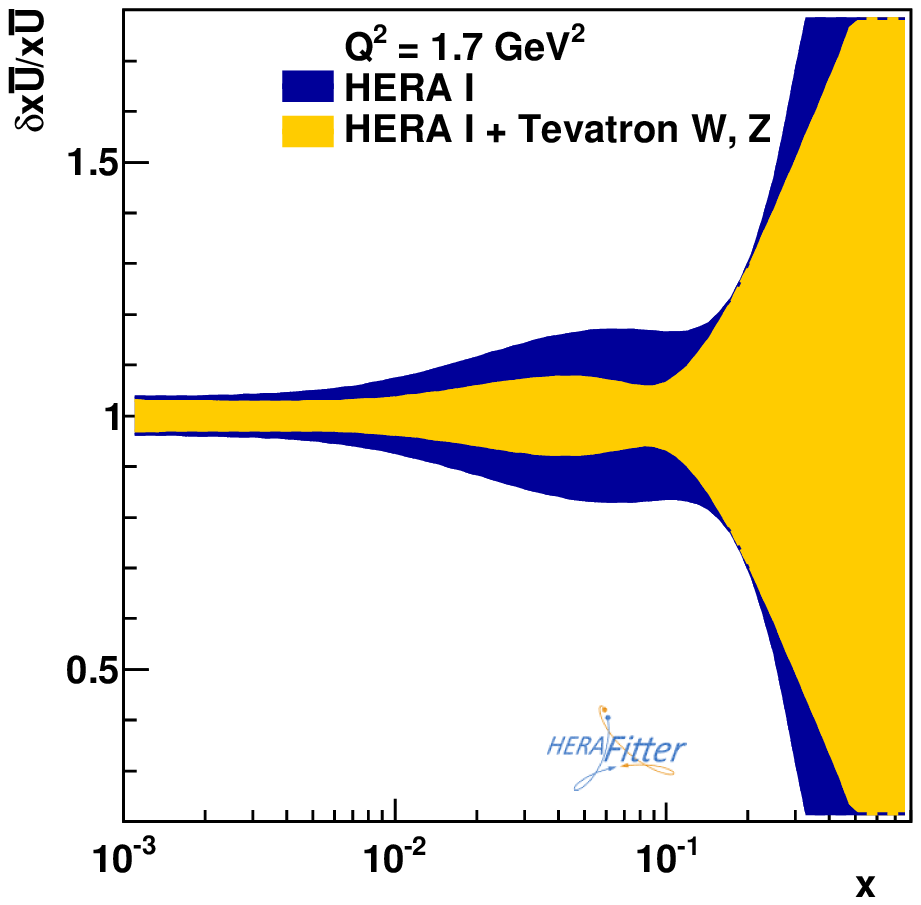}}
    \subfigure[]{\includegraphics[width=0.49\textwidth]{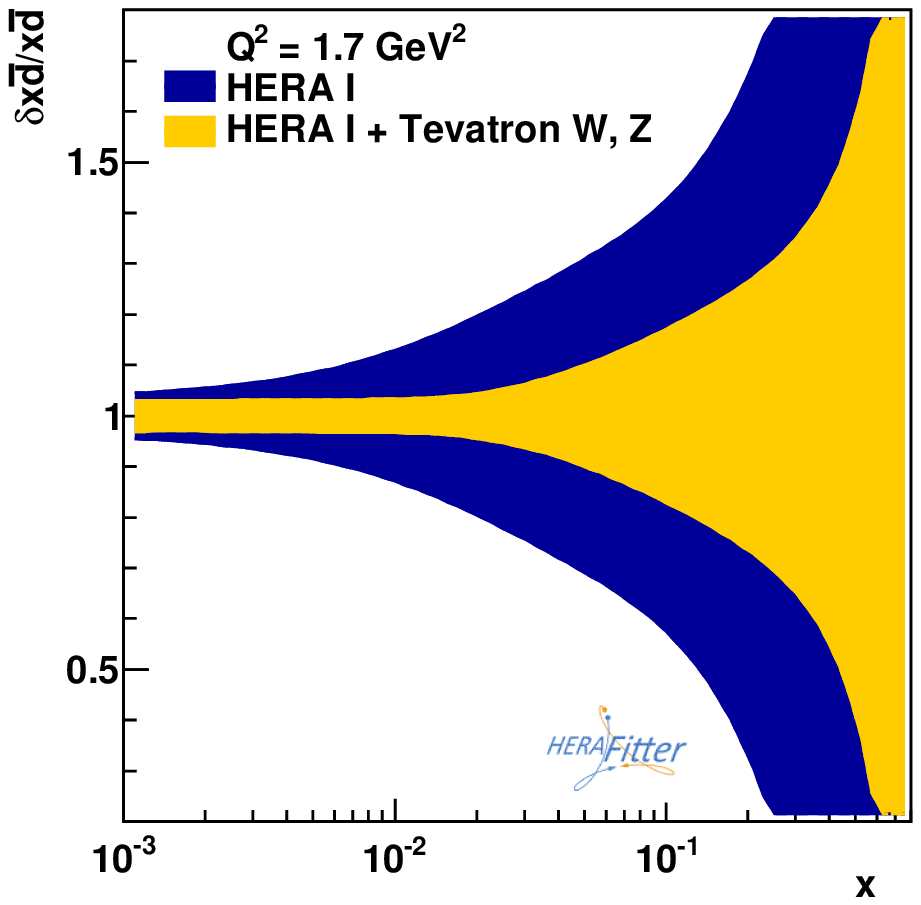}}
  \end{center}
  \caption{\label{fig:pdfratio} Relative PDF uncertainties 
   at the starting scale $Q^2 =1.7$~GeV$^2$ as a function of
   Bjorken-$x$ for (a) $u_v$, (b) $d_v$, (c) $\bar{u}$, and (d)
   $\bar{d}$, determined with a fit to the HERA I data (blue), and with a fit
   to the HERA I and Tevatron $W$- and $Z$-boson data (yellow).
}
\end{figure*}

The central value and the uncertainties of the PDFs are evaluated with
MC replicas~\cite{Forte:2002fg}: the data points are smeared using
Gaussian distributions, according to their experimental uncertainties,
and the PDF fit is repeated 1000 times, using different random seeds
for the smearing. The central PDFs are calculated as the average of the replicas
and the PDF uncertainties are calculated from their standard
deviation. Figure~\ref{fig:pdffit} shows the comparison
of the PDFs extracted with the MC-replica method by fitting the HERA I
data, and by fitting the HERA I and Tevatron $W$- and $Z$-boson
data. Figure~\ref{fig:pdfratio} shows the comparison of the relative
uncertainty of the two PDFs. A significant reduction of the PDF
uncertainties is observed in the fit which includes the Tevatron $W$-
and $Z$-boson measurements, in particular for the valence quarks and
$\bar{d}$ quarks.

A fit of the HERA I and Tevatron $W$- and $Z$-boson data with the same
settings, but with a correlation model in which trigger and
identification uncertainties are treated as correlated bin-to-bin,
yields very similar central PDFs and PDF uncertainties [not
shown]. The $\chi^2$ of all the data sets are also very similar, except
for the $\chi^2$ of the CDF $W$-boson asymmetry measurements, which is
about twice as large.

\begin{figure*}
    \begin{center}
    \subfigure[]{\includegraphics[width=0.49\textwidth]{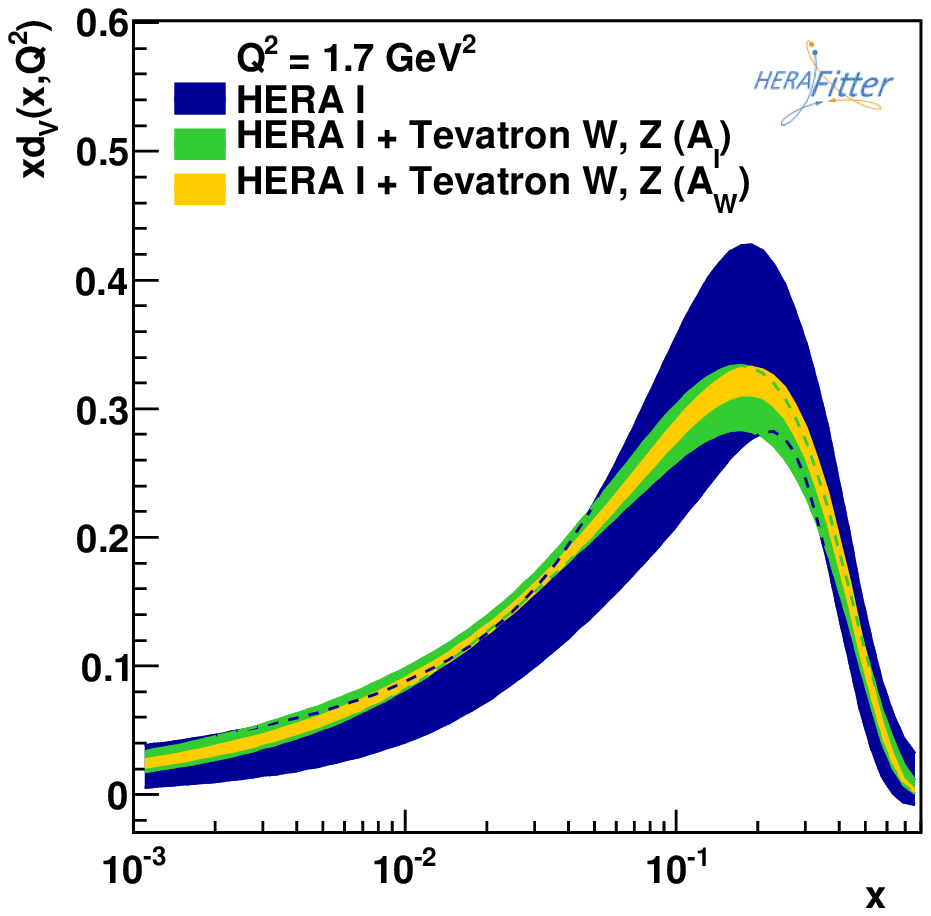}}
    \subfigure[]{\includegraphics[width=0.49\textwidth]{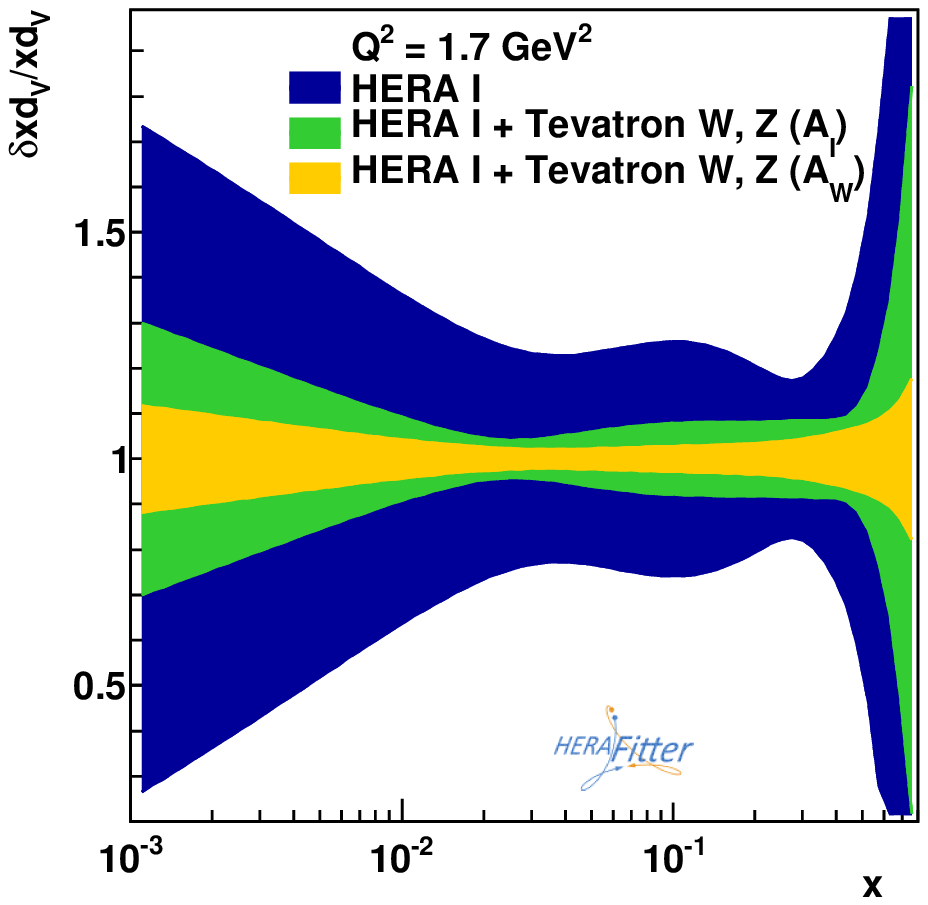}}
  \end{center}
  \caption{\label{fig:wleptfit} (a) $d$-valence PDF at the scale $Q^2 =
    1.7$~GeV$^2$ as a function of Bjorken-$x$ and (b) $d$-valence relative
    PDF uncertainties, determined with a fit to the HERA I data
    (blue), with a fit to the HERA I and Tevatron $W$-boson asymmetry
    and $Z$-boson data (yellow), and with a fit to the HERA I and
    Tevatron $W$-boson lepton asymmetry and $Z$-boson data (green).}
\end{figure*}

The $W$-boson charge asymmetries rely on the reconstruction of the
$W$-boson rapidity, which is measured assuming a fixed $W$-boson mass, and
inferring the unmeasured longitudinal momentum of the neutrino on a
statistical basis~\cite{Aaltonen:2009ta,Abazov:2013dsa}. The
reconstruction of the $W$-boson rapidity introduces a model dependence
in the measurement.
To study the possible bias due to the $W$-boson rapidity
reconstruction, an alternative fit is performed in which the
$W$-boson charge asymmetries measured by CDF and D0 are excluded, and
the latest D0 measurement of the electron asymmetry is included.
The $\chi^2_{\textrm{min}}$ / dof and the partial $\chi^2$ of the fit are shown in
Tab.~\ref{tab:wleptfit}. Also for this fit the partial $\chi^2$ 
of each of the Tevatron and HERA I data set is close
to unity. The $d_v$ PDF determined from the fit is
shown in Fig.~\ref{fig:wleptfit}, and compared to the nominal fit. The
fit to the lepton asymmetry data yields very compatible results, but the
uncertainties on the $d_v$ PDF are up to twice as large.

The impact of posterior inclusion of 
the Tevatron $W$- and $Z$-boson measurements on the PDF 
uncertainties as estimated by
CT10nlo, MMHT2014, and NNPDF3.0  is assessed by profiling and
reweighting. For consistency with the other PDFs, the uncertainties of
the CT10nlo PDFs are scaled to 68\% confidence interval by applying a factor of 1.645.
The three PDF sets already include the CDF and D0 $Z$-boson differential cross
sections as a function of rapidity, and the MMHT2014 fit also includes
the D0 muon charge asymmetry in $W \to \mu \nu$ decays and the CDF $W$
charge asymmetry in the $W \to e \nu$ decay channel. Only the
measurements that are not included in each of the PDF sets are
considered for the corresponding profiling or reweighting.
The compatibility of the Tevatron data with the
CT10nlo, MMHT2014 and NNPDF3.0 sets is tested by evaluating the
$\chi^2$ function of Eq.~(\ref{eq:chi2prof}), accounting for asymmetric
PDF uncertainties according to Eq.~(\ref{eq:iter}).
To perform this calculation for the NNPDF3.0 set, the covariance matrix 
for the predictions is decomposed using the eigenvector representation.
Table~\ref{tab:chi2profiling} shows the compatibility between the
Tevatron measurements and the above PDF sets, together with the partial
$\chi^2$ of each data set. The partial $\chi^2$ per number of points
of each of the Tevatron data set, and the total $\chi^2$ / dof, are
close to unity for all the PDFs, when the $\chi^2$ evaluation includes
the PDF uncertainties. The quality of the agreement significantly
deteriorates if the $\chi^2$ evaluation neglects the PDF
uncertainty. This effect is more pronounced for the CT10nlo and
NNPDF3.0 sets which include fewer data from the Tevatron. This
indicates the significant constraining power of the Tevatron data.
 
\begin{figure*}
  \begin{center}
    \subfigure[]{\includegraphics[width=0.49\textwidth]{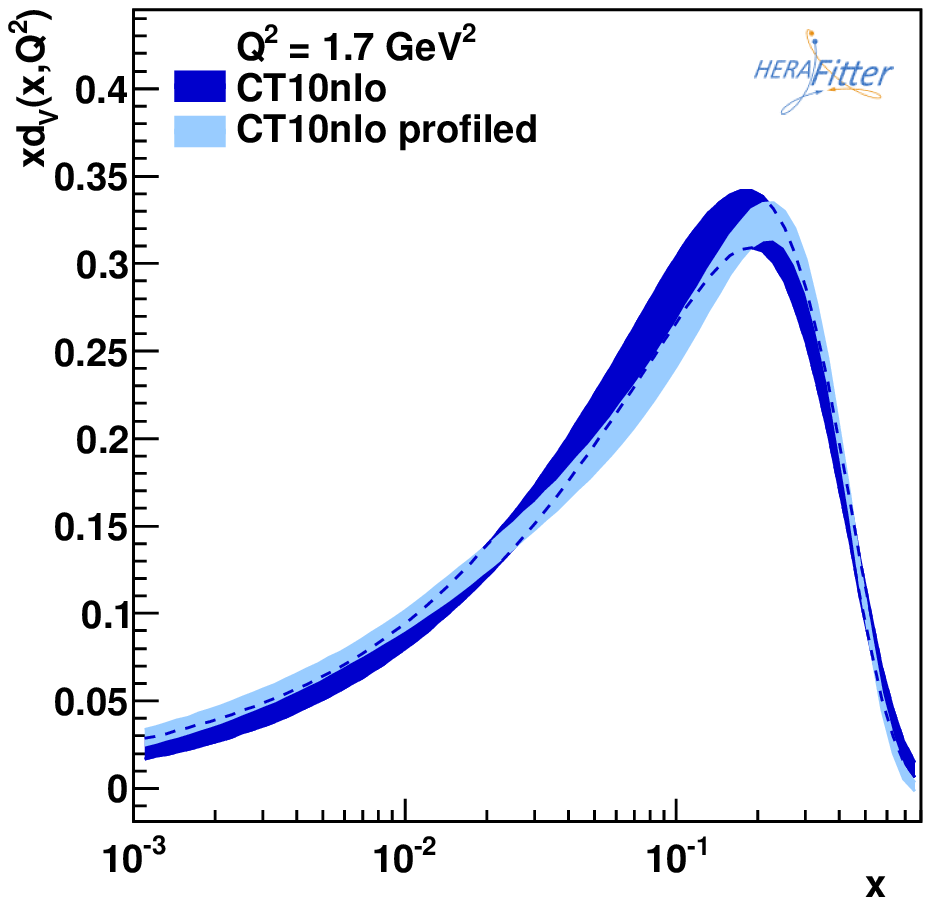}}
    \subfigure[]{\includegraphics[width=0.49\textwidth]{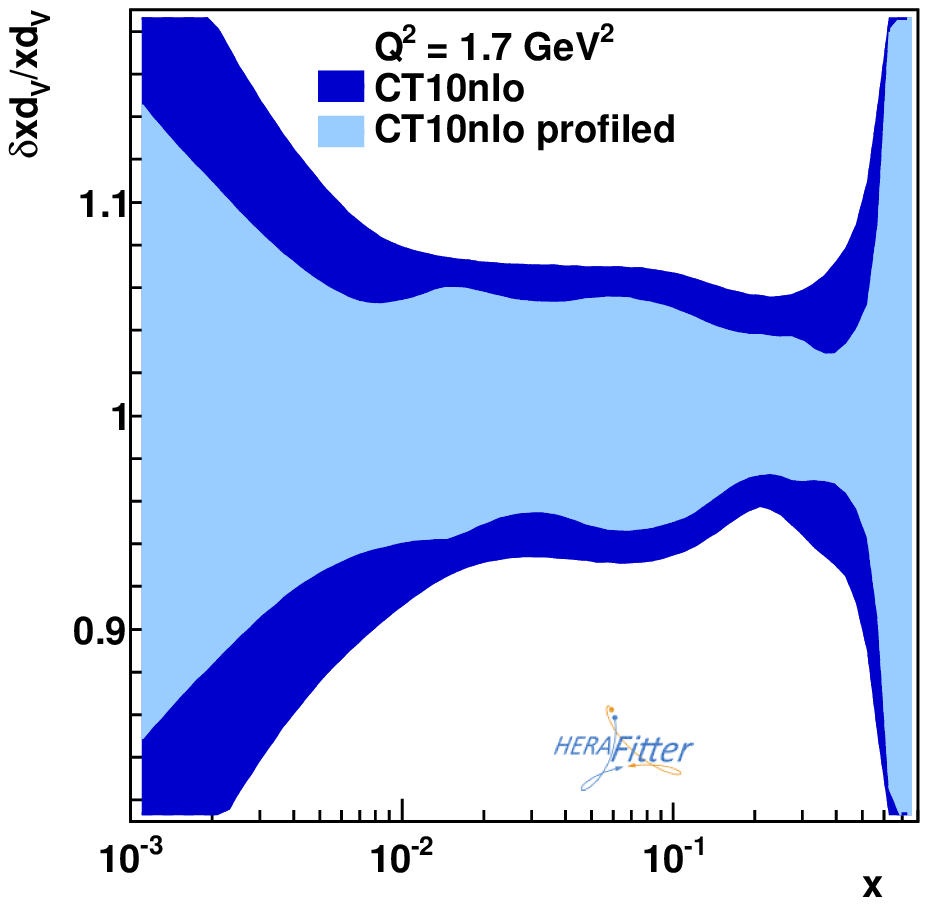}}\\
    \subfigure[]{\includegraphics[width=0.49\textwidth]{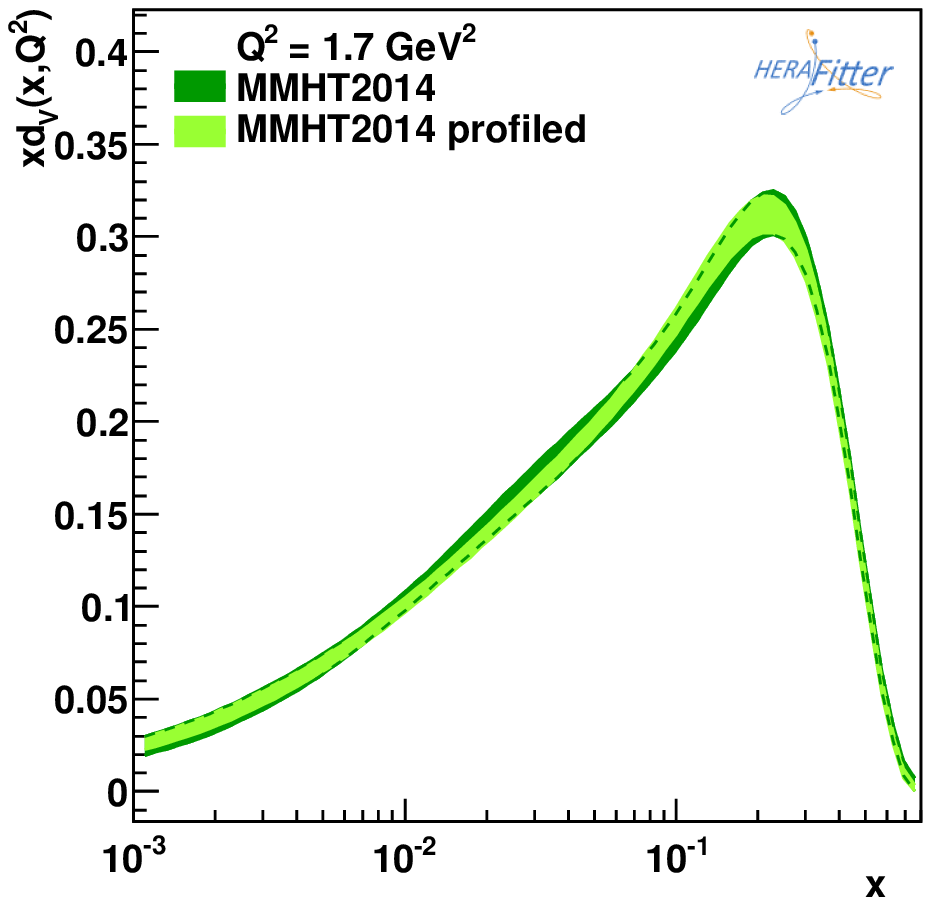}}
    \subfigure[]{\includegraphics[width=0.49\textwidth]{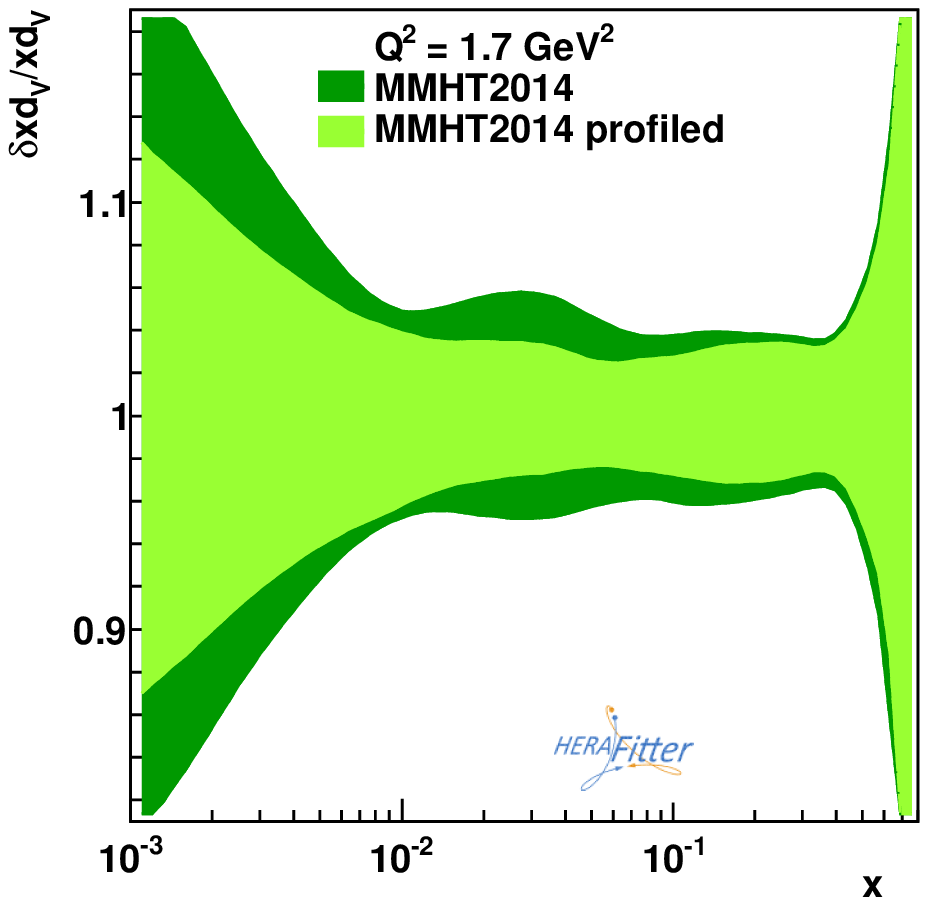}}\\
    \caption{$d$-valence PDF at the scale $Q^2 = 1.7$~GeV$^2$ as a function of Bjorken-$x$ before
    and after profiling for the (a) CT10nlo and (c) MMHT2014 PDFs and
    the corresponding relative uncertainties for (b) CT10nlo and (c) MMHT2014.
\label{fig:pdfprofiled}}
  \end{center}
\end{figure*}

The CT10nlo and MMHT2014 PDFs are profiled according to
Eq.~(\ref{eq:prof}). The results of the profiling on the $d$-valence
PDFs, and on their relative uncertainties, are shown in
Fig.~\ref{fig:pdfprofiled}. The profiling affects the shape of the
distribution  more for the CT10nlo when compared to MMHT2014
set. Significant reduction of the uncertainties is observed for both
sets, in particular in the low- and medium-$x$ range. The NNPDF3.0 PDFs
are reweighted to the Tevatron data. The number of effective replicas
remaining  after reweighting, $N_\mathrm{eff}$, is only $1$ and hence
the resulting PDFs are not shown.

\begin{figure*}
  \begin{center}
    \subfigure[]{\includegraphics[width=0.49\textwidth]{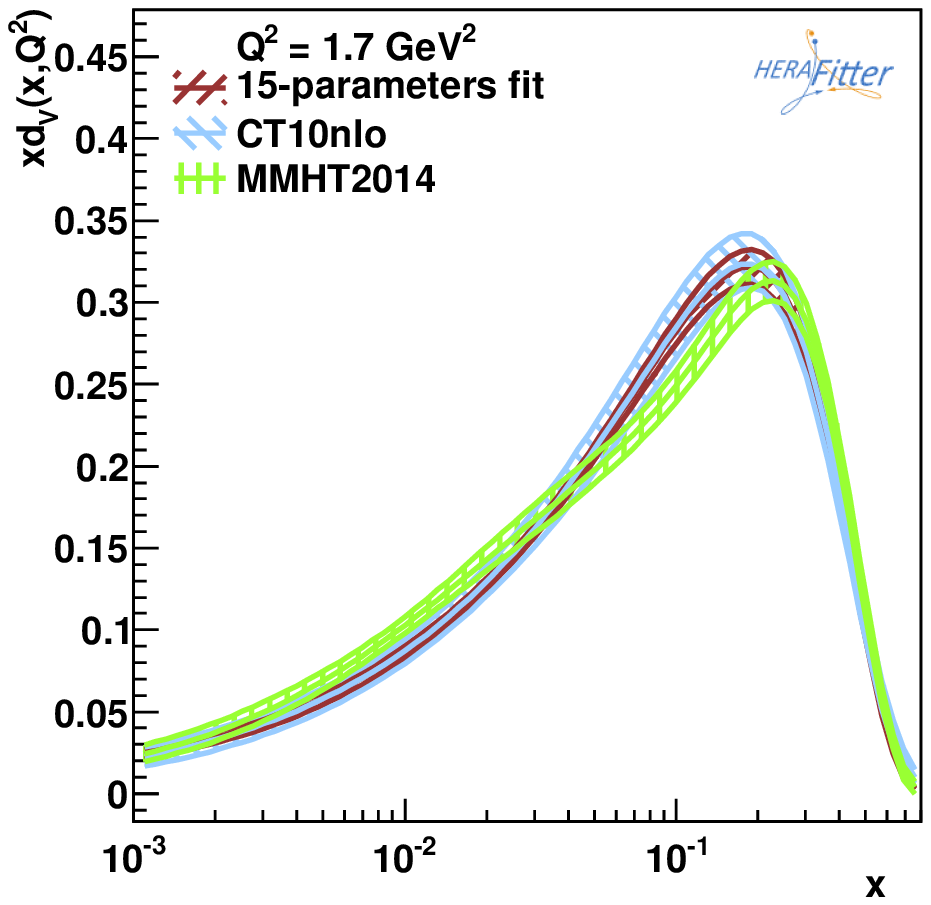}}
    \subfigure[]{\includegraphics[width=0.49\textwidth]{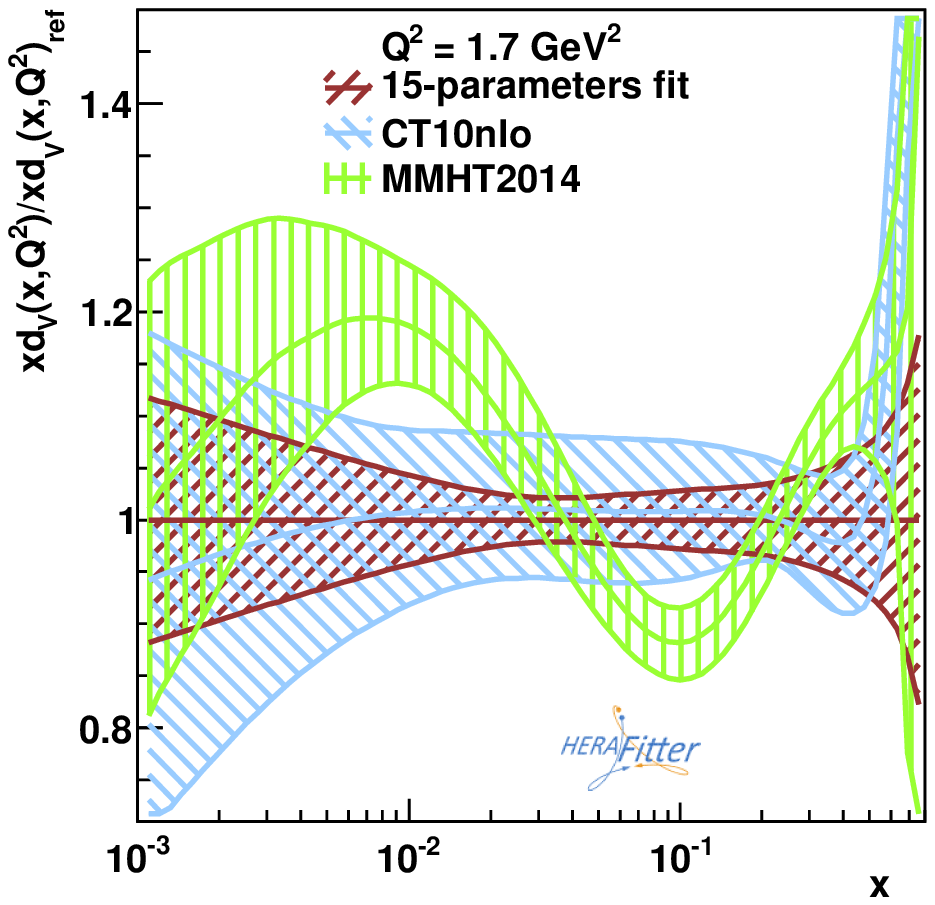}}\\
    \caption{\label{fig:summarydv_orig} (a) $d$-valence PDF at the
      scale $Q^2 = 1.7$~GeV$^2$ as a function of Bjorken-$x$ determined from a fit
    to the HERA I and Tevatron $W$- and $Z$-boson data, and from the CT10nlo and
    MMHT2014 PDFs; (b) ratio of $d$-valence PDFs central values and uncertainties
    with respect to the $d$-valence PDF determined from a fit the HERA I and
    Tevatron $W$- and $Z$-boson data.
}
  \end{center}
\end{figure*}

\begin{figure*}
  \begin{center}
    \subfigure[]{\includegraphics[width=0.49\textwidth]{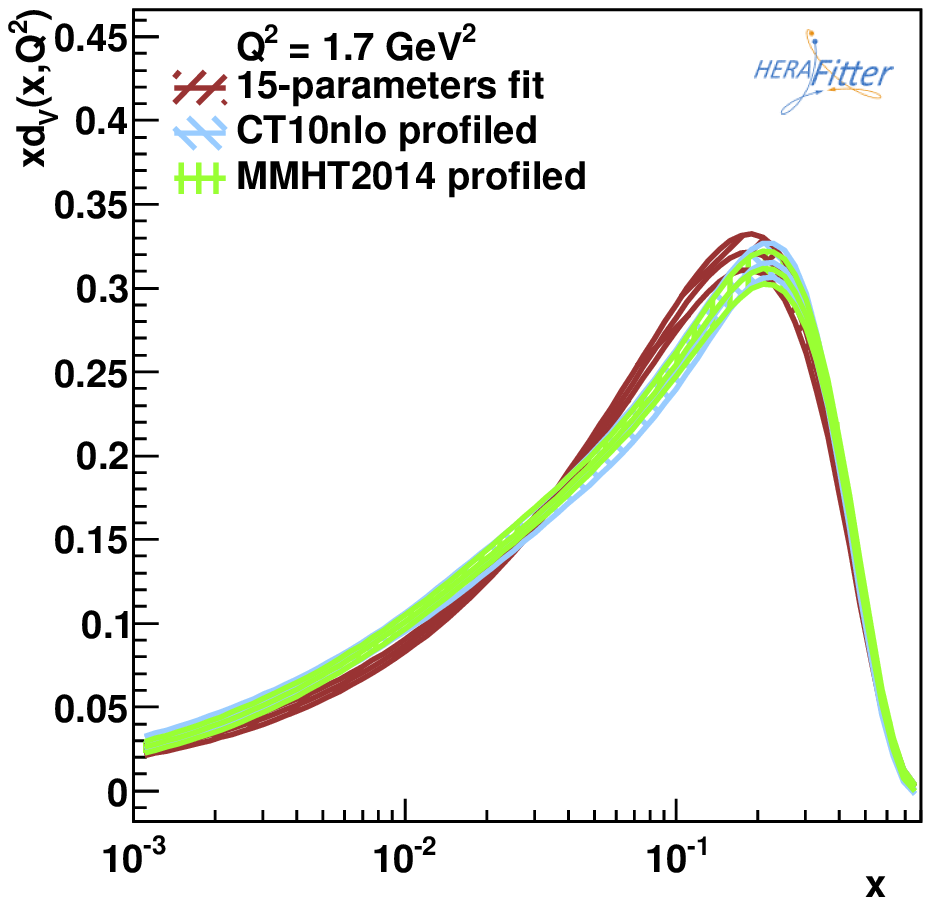}}
    \subfigure[]{\includegraphics[width=0.49\textwidth]{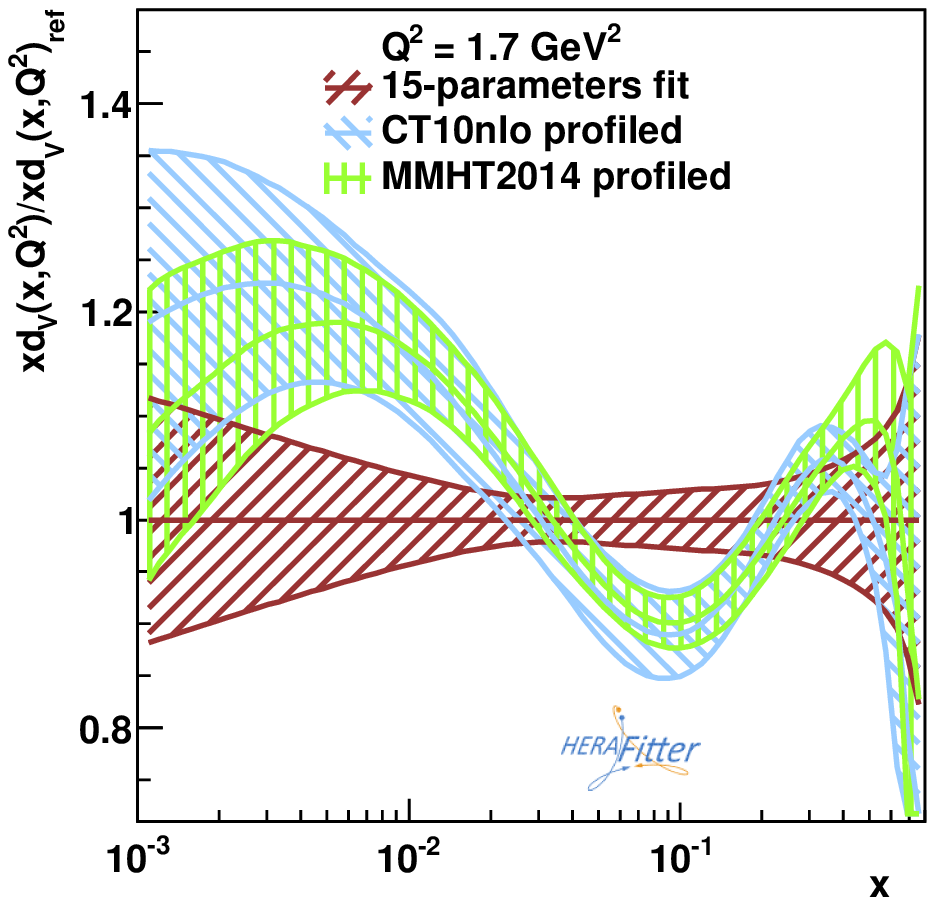}}\\
    \caption{\label{fig:summarydv}  (a) $d$-valence PDF at the
      scale $Q^2 = 1.7$~GeV$^2$ as a function of Bjorken-$x$ determined from a fit
    to the HERA I and Tevatron $W$- and $Z$-boson data, and from the
    profiled CT10nlo and
    MMHT2014 PDFs; (b) ratio of $d$-valence PDFs central values and uncertainties
    with respect to the $d$-valence PDF determined from a fit the HERA I and
    Tevatron $W$- and $Z$-boson data.
}
  \end{center}
\end{figure*}

The original and profiled $d$-valence PDFs, and the result of the fit
to the HERA I and Tevatron $W$- and $Z$-boson data, are compared in
Figs.~\ref{fig:summarydv_orig} and~\ref{fig:summarydv},
respectively. The profiling using Tevatron data improves agreement of
the $d$-valence distribution between the MMHT2014 and CT10nlo PDF
sets.

\section{Summary}
\label{sec:conclusions}
The \HERAFitter framework is used to perform a QCD analysis
of the DIS data from HERA, together with $W$- and $Z$-boson production
measurements performed at the Tevatron collider in Run II. The
correlation model of the systematic uncertainties of the Tevatron data
is investigated, and a modification is proposed which accounts for the
statistical nature of some of the systematic uncertainties. The
Tevatron and HERA data are well described by a NLO fit with $15$
free parameters, with a new parametrisation of PDFs which adds to the
basic form a combination of linear and exponential terms.
The impact of the Tevatron $W$- and $Z$-boson measurements is assessed
by comparing PDF uncertainties from a fit to the HERA data alone, and
a fit to the HERA and Tevatron data. 
A significant reduction of the uncertainties is observed in the latter case,
for the valence quarks and $\bar{d}$ quarks in particular.

The Tevatron measurements are also compared to predictions evaluated
with modern PDF sets, and the impact of the data on the PDFs is
assessed using profiling and reweighting techniques.
The profiling techniques take
into account asymmetric PDF
uncertainties. A good agreement between measurements and predictions
is observed, if the PDF uncertainties of the predictions are taken
into account.
After the inclusion of the Tevatron data, the PDF uncertainties on the
$d$-valence quarks are significantly reduced, especially for the PDF
sets which include only the $Z$-boson data from the Tevatron, and the
agreement between the various PDF sets is improved. These findings
highlight the importance of the Tevatron $W$- and $Z$-boson production
data to constrain $d$-quark and valence PDFs, and suggest that the data
should be used in the future global PDF analyses.
All the supporting material to allow fits of the Tevatron data,
including the updated correlation model and the grid files for fast theory
calculations, are publicly available on the web page of the \HERAFitter
project.
\begin{acknowledgements}
~\\
We thank Juan Rojo for useful comments on the manuscript, and Hang Yin
for fruitful discussions on the experimental uncertainties of the D0 measurements.
We are grateful to the DESY IT department for their support of the 
\HERAFitter developers. 
This work is supported in part by  Helmholtz Gemeinschaft under
contract VH-HA-101, BMBF-JINR cooperation and Heisenberg-Landau programs
as well as by the Polish NSC project DEC-2011/03/B/ST2/00220.
\end{acknowledgements}
\FloatBarrier
\bibliographystyle{utphys_mod} 
\bibliography{main}

\begin{table*}
  \begin{center}
    \caption{\label{tab:tevwzdata} Summary of the Tevatron $W$- and $Z$-boson measurements. 
      For each measurement the observable, the experiment, the
      integrated luminosity, the phase-space definition, the inclusion
      in the nominal fit, and the corresponding Ref. are shown.}
    \begin{tabular}{lcccccc}
      \toprule
      Observable       & Experiment & Integrated     & Kinematic    &   Used in the  & Ref.\\
                       &            & luminosity     & requirements &    nominal fit & \\
      \midrule
      $d\sigma(Z)/dy$                                  & D0  & 0.4 fb$^{-1}$& $71 < m_{ee} < 111$~GeV                  & yes & \cite{Abazov:2007jy}\\
      $d\sigma(Z)/dy$                                  & CDF & 2.1 fb$^{-1}$& $66 < m_{ee} < 116$~GeV                  & yes & \cite{Aaltonen:2010zza}\\
      muon charge asymmetry in $W \to \mu \nu$         & D0  & 7.3 fb$^{-1}$& $p_T^{\mu} > 25$~GeV, $p_T^{\nu} > 25$~GeV & yes & \cite{Abazov:2013rja}\\
      electron charge asymmetry in $W \to e \nu$       & D0  & 9.7 fb$^{-1}$& $E_T^{e} > 25$~GeV, $p_T^{\nu} > 25$~GeV   & no  & \cite{D0:2014kma}\\
      $W$ charge asymmetry in $W \to e \nu$            & CDF & 1.0 fb$^{-1}$&  none                                   & yes & \cite{Aaltonen:2009ta}\\
      $W$ charge asymmetry in  $W \to e \nu$           & D0  & 9.7 fb$^{-1}$& $E_T^{e} > 25$~GeV, $p_T^{\nu} > 25$~GeV   & yes  & \cite{Abazov:2013dsa}\\
      \bottomrule
    \end{tabular}
  \end{center}
\end{table*}

\begin{table*}
\begin{center}
\caption{\label{tab:parscan} Results of the parametrisation study. For
  each additional free parameter $D$, $E$, and $F$, of the $d_v$,
  $u_v$, gluon, $\bar{u}$, and $\bar{d}$ PDF, the reduction of
  $\chi^2_{\textrm{min}}$ of a fit to the Tevatron and HERA I data,
  $\Delta \chi^2_{\textrm{min}}$, is shown. For each of the fit with
  $n$ free parameters, with $n=10,11,12,13,14$, the largest $\Delta
  \chi^2_{\textrm{min}}$ is shown in bold, and the corresponding
  parameter is added as a free parameter for the $n+1$-parameters fit.
  The fits which leads to negative high-x PDFs are shown in parenthesis, and
  are not considered in the parametrisation study.
}
\begin{tabular}{ccccccc}
\toprule
$n$     &   10         &      11      &      12     &       13      &       14\\
$\chi^2_{\textrm{min}}/\textrm{dof}$ & 714/633 &  654/632      &     619/631    &       610/630    &     607/629\\
\midrule
Free parameter & \multicolumn{5}{c}{$\Delta \chi^2_{\textrm{min}}$} \\
\midrule
\multicolumn{1}{l}{$d_v$} \\
D 	&	39	&	-	&	-	&	-	&	-\\
E	&	40	&	-	&	-	&	-	&	-\\
F	&\textbf{41}	&	-	&	-	&	-	&	-  \\
\midrule
\multicolumn{1}{l}{$u_v$} \\
D	&	$<1$	&	41	&	-	&	-	&	-\\
E	&	1	&	38	&	-	&	-	&	-\\
F	&	$<1$	&\textbf{44}	&	-	&	-	&	-  \\
\midrule
\multicolumn{1}{l}{gluon} \\
D	&	25	&	16	&\textbf{9.8}	&	-	&	-  \\
E	&	9	&	8	&	5.4	&	$<0.1$	&	0.3\\
F	&	35	&	19	&	7.2	&	-	&	-\\
\midrule
\multicolumn{1}{l}{$\bar{d}$} \\
D	&	2	&	$<1$	&	1.9	&\textbf{3.0}	&	-  \\
E	&	(44)	&	(3)	&	(3.5)	&	(3.4)	&	(1.9)\\
F	&	5	&	1	&	$<0.1$	&	0.1	&	-\\
\midrule
\multicolumn{1}{l}{$\bar{u}$}\\
D	&	(44)	&	3	&	1.6	&	0.6	&\textbf{1.5}\\
E	&	(49)	&	(8)	&	(1.9)	&	(1.3)	&	(1.5)\\
F	&	13	&	2	&	1.3	&	0.7	&	1.0\\
\bottomrule
\end{tabular}
\end{center}
\end{table*}

\begin{table*}
  \begin{center}
    \caption{\label{tab:chi2fit} Results of a 15-parameters fit to the
    HERA I data and to the HERA I and Tevatron $W$- and $Z$-boson
    data. The contribution to the total $\chi^2_{\textrm{min}}$ of
    each data set and the corresponding number of points are shown.}
    \begin{tabular}{lcc}
      \toprule
      & HERA I   & HERA I + Tevatron W, Z \\
      Data set       & $\chi^2$ / number of points & $\chi^2$ / number of points \\
      \midrule
      NC DIS cross sections H1-ZEUS combined $e^-p$. & 112 / 145& 106 / 145  \\
      NC DIS cross sections H1-ZEUS combined $e^+p$. & 326 / 337& 334 / 337  \\
      CC DIS cross sections H1-ZEUS combined $e^-p$. & 20 / 34& 19 / 34  \\
      CC DIS cross sections H1-ZEUS combined $e^+p$. & 27 / 34& 33 / 34  \\
      HERA I correlated $\chi^2$  & 21& 23  \\
      \midrule
      D0 $d\sigma(Z)/dy$  & - & 23 / 28  \\
      CDF $d\sigma(Z)/dy$  & - & 33 / 28  \\
      D0 muon charge asymmetry in $W \to \mu \nu$ & - & 12 / 10  \\
      CDF $W$ charge asymmetry in $W \to e \nu$ & - & 15 / 13  \\
      D0 $W$ charge asymmetry in  $W \to e \nu$ & - & 16 / 14  \\
      \midrule
      Total $\chi^2_{\textrm{min}}$ / dof  & 505 / 535& 615 / 628  \\
      \bottomrule
    \end{tabular}
  \end{center}
\end{table*}

\begin{table*}
  \begin{center}
    \caption{\label{tab:chi2profiling} Comparison between the Tevatron
    $W$-boson measurements and the CT10nlo, MMHT2014, and NNPDF3.0
    PDFs. The partial $\chi^2$ of each data set, the total $\chi^2$,
    and the total $\chi^2$ without PDFs uncertainties are shown.}
    \begin{tabular}{lccc}
      \toprule
      PDF set     & CT10nlo   & MMHT2014   & NNPDF3.0  \\
             & $\chi^2$ / number of points & $\chi^2$ / number of points & $\chi^2$ / number of points \\
      \midrule
      D0 muon charge asymmetry in $W \to \mu \nu$  & 13 / 10 &  -     & 12 / 10  \\
      CDF $W$ charge asymmetry in $W \to e \nu$    & 15 / 13 &  -     & 17 / 13  \\
      D0 $W$ charge asymmetry in  $W \to e \nu$    & 15  / 14 &  10/14  & 3  / 14  \\
      PDF correlated $\chi^2$                      & 4       & 3      & 6 \\
      \midrule
        Total $\chi^2$ / dof                       & 47 / 37 & 14 / 14  & 39 / 37  \\
      \midrule
        Total $\chi^2$ / dof without PDFs uncertainties & 368/37 & 42/14 & 846 / 37  \\
      \bottomrule
    \end{tabular}
  \end{center}
\end{table*}

\begin{table*}
  \begin{center}
    \caption{\label{tab:wleptfit} Results of a 15-parameters fit to
      the HERA I and Tevatron $W$-boson lepton asymmetry and $Z$-boson
      data. The contribution to the total $\chi^2_{\textrm{min}}$ of
      each data set and the corresponding number of points are shown.}
    \begin{tabular}{lc}
      \toprule
      & HERA I + Tevatron W-lepton, Z \\
      Data set       & $\chi^2$ / number of points \\
      \midrule
      NC DIS cross sections H1-ZEUS combined $e^-p$. &  107 /  145\\
      NC DIS cross sections H1-ZEUS combined $e^+p$. &  334 /  337\\
      CC DIS cross sections H1-ZEUS combined $e^-p$. &   20 /   34\\
      CC DIS cross sections H1-ZEUS combined $e^+p$. &   32 /   34\\
      HERA I correlated $\chi^2$  & 22 \\
      \midrule
      D0 $d\sigma(Z)/dy$  & 23/ 28\\
      CDF $d\sigma(Z)/dy$ & 32/ 28 \\
      D0 muon charge asymmetry in $W \to \mu \nu$ & 13 / 10 \\
      D0 electron charge asymmetry in $W \to e \nu$ & 19 / 13 \\
      \midrule
      Total $\chi^2_{\textrm{min}}$ / dof  & 603 / 614  \\
      \bottomrule
    \end{tabular}
  \end{center}
\end{table*}

\end{document}